\documentclass[12pt,fleqn]{uofsthesis-cs}
\usepackage[utf8x]{luainputenc}
\usepackage{etoolbox}
\usepackage{float}
\usepackage{xcolor}
\definecolor{lightblue}{HTML}{86afdb}
\definecolor{lightgreen}{HTML}{87ba6e}
\usepackage{relsize}
\usepackage{caption}
\usepackage{subcaption}
\usepackage{amsmath,amssymb,slashed,simplewick,bbold}
\usepackage{graphicx}
\usepackage[noblocks]{authblk}
\usepackage{setspace}
\usepackage{braket}
\usepackage{array}
\usepackage{hyperref}
\usepackage{multirow}
\usepackage{soul}
\usepackage{tikz}
\usepackage[numbers, sort, compress]{natbib}
\usepackage[compat=1.1.0]{tikz-feynman}
\usepackage{indentfirst}

\usepackage{multirow}
\usepackage{xcolor,colortbl}
\usepackage{hhline}
\usepackage{array}
\newcolumntype{?}{!{\vrule width 1pt}}
\makeatletter
\def\thickhline{%
  \noalign{\ifnum0=`}\fi\hrule \@height \thickarrayrulewidth \futurelet
   \reserved@a\@xthickhline}
\def\@xthickhline{\ifx\reserved@a\thickhline
               \vskip\doublerulesep
               \vskip-\thickarrayrulewidth
             \fi
      \ifnum0=`{\fi}}
\makeatother

\newlength{\thickarrayrulewidth}
\title{Heavy-Light and Doubly-Strange Diquark Spectrum from QCD Laplace Sum-Rules and Diagrammatic Renormalization Methods}
\author{Thamirys de Oliveira}
\degree{\PhD}%{\MSc}          
\defencedate{August 2023}
\department{Physics and Engineering Physics}
\academicunit{Department}
\department{Physics and Engineering Physics}
\ptuaddress{Department of Physics \& Engineering Physics\\
University of Saskatchewan\\
116 Science Place, Rm 163\\
Saskatoon, SK S7N 5E2\\
Canada
}

\abstract{
In addition to the conventional $q\bar{q}$ and $qqq$ bound states of quarks $q$, exotic hadrons, such as the four-quark $q\bar{q}q\bar{q}$ and the five-quark $qqqq\bar{q}$ bound states, can be constructed by assuming colour confinement in strong interactions. In particular, one of the possible internal structure for a four-quark state is the tetraquark $[qq][\bar{q}\bar{q}]$, where diquarks $[qq]$ and antidiquarks $[\bar{q}\bar{q}]$ are bound due to the colour force. In this thesis, heavy-light $[Qq]$, where $Q\in\{c, b\}$ (charm, bottom) and $q\in\{u,d,s\}$ (up, down, strange), and doubly-strange $[ss]$ diquarks are examined using QCD Laplace sum-rules. The diquark two-point correlation function is renormalized using diagrammatic renormalization methods for QCD correlation functions that are developed and compared to the conventional renormalization method. It is shown that the mixing of composite operators induced by the conventional renormalization approach is avoided by the diagrammatic renormalization method. The strange quark condensate parameter $\kappa=\langle\bar{s}s\rangle/\langle\bar{n}n\rangle$, where $n\in\{u,d\}$, is shown to have an important role on the $[Qs]$ and $[Qn]$ diquark mass splitting.
}

\acknowledgements{I would like to thank my supervisor Dr.~Tom Steele for his support and advice, and for the inspiring weekly coffees. I would also like to thank collaborators and research group members: Dr.~Derek Harnett, Dr.~Robin Kleiv, Dr.~Alex Palameta, Dr.~Jason Ho, MSc.~Siyuan Li, and MSc.~Barbara Cid Mora. Special thanks to my family for their continued encouragement.
}

\dedication{ 
In memory of Camila de Oliveira.}

\loa{
\abbrev{1PI}{One-Particle Irreducible}
\abbrev{LHS}{Left-hand Side}
\abbrev{LO}{Leading Order}
\abbrev{MS-scheme}{Minimal Subtraction Scheme}
\abbrev{$\overline{\text{MS}}$-scheme}{Modified Minimal Subtraction Scheme}
\abbrev{NLO}{Next-to-leading Order}
\abbrev{OPE}{Operator Product Expansion}
\abbrev{RHS}{Right-hand Side}
\abbrev{SI units}{International System of Units}
\abbrev{SM}{Standard Model}
\abbrev{VEV}{Vacuum Expectation Value}
\abbrev{QCD}{Quantum Chromodynamics}
\abbrev{QCDSR}{QCD Sum-Rules}
\abbrev{QED}{Quantum Electrodynamics}
\abbrev{QFT}{Quantum Field Theory}
}

\begin{document}

\maketitle

\frontmatter
\chapter{Introduction}
\label{Introduction}

\section{Motivation}
\label{Sec:Motivation}

Since 2003, with the discovery of the $X(3872)$ \cite{X3872:PhysRevLett.91.262001} state, dozens of mesons with properties that do not fit as conventional $q\bar{q}$ bound states have been measured (see e.g., Refs.~\cite{Swanson:2006st,Godfrey:2008nc,Olsen:RevModPhys.90.015003,Brambilla:2019esw} for reviews). Recently, for instance, the LHCb collaboration observed a doubly charged state, the $T^a_{c\bar{s}0}(2900)^{++}$ \cite{LHCb:2022xob}. Since it is not possible to form a doubly charged state from a $q\bar{q}$ pair, the $T^a_{c\bar{s}0}(2900)^{++}$ is a tetraquark state candidate. Tetraquarks are formed by a coloured diquark $[qq]$ and a coloured antidiquark $[\bar{q}\bar{q}]$ interacting in a similar way that a $q$ and an $\bar{q}$ interact to form a colour singlet. Diquark masses appear as an important parameter in tetraquark models (see, e.g., Ref.~\cite{Maiani:2004vq}), where the diquark mass is determined, for example, within the model itself. Therefore, it is important to give QCD support to the constituent diquark masses used in these models.

In this thesis, QCD sum-rules are used to connect two-point correlation function of composite operators with a hadronic spectral function, allowing, for example, the prediction of diquark masses. In this process, the correlation function needs to be renormalized in order to remove all non-local divergences. In the conventional renormalization of the correlation function of composite operators the complexity of the renormalization process increases as the mass dimension of the operator increases due to the mixing of operators. In Chapter \ref{chapter:Paper_1}, the diagrammatic renormalization method is studied and applied in the context of QCD sum-rules, and it is shown that operator mixing is avoided with the diagrammatic renormalization, increasing the efficiency of the renormalization process.

In Chapter \ref{chapter:Paper_2}, heavy-light $[Qq]$, where $Q\in\{c,b\}$ and $q\in\{u,d,s\}$, and doubly-strange $[ss]$ diquark two-point correlation functions are renormalized using the diagrammatic renormalization method. Diquark masses are extracted in the context of QCD Laplace sum-rules when a stable sum-rule is found. The mass splitting between the heavy-strange $[Qs]$ and the heavy-non-strange $[Qn]$ diquarks, where $n\in\{u,d\}$, is found to be significantly affected by the uncertainty in the strange quark condensate parameter $\kappa=\langle\bar{s}s\rangle/\langle\bar{n}n\rangle$. The theoretical uncertainty from all other input parameters are reduced by an analysis methodology developed in Chapter \ref{chapter:Paper_2}. In addition, this thesis is working in Minkowsi (Lorentzian) space-time with a metric convention where time-like four vectors have a positive scalar product, and the Einstein summation convention will be used implicitly (see Appendix \ref{Appendix_A} for details on these and other conventions).
\section{Conventional Quark Model}
\label{Sec:CQM}

\begin{figure}[ht]
	\centering
	\includegraphics{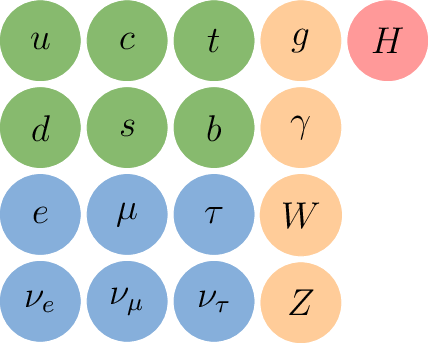}
	\caption{Elementary particles in the Standard Model. The first three columns are the first, second, and third electroweak \textit{generations}, respectively, of quarks (in green) and leptons (in blue).}
	\label{fig:SM}
\end{figure}

The Standard Model (SM) describes the electromagnetic, weak, and strong interactions of nature\footnote{Electromagnetism and weak interactions are two manifestations of the unified electroweak theory.}. These interactions are mediated by the exchange of spin-1 gauge bosons, also known as force carries, between particles. Electromagnetic interactions are mediated by the exchange of photons ($\gamma$), weak interactions are mediated by the exchange of massive $W^{\pm}$ and $Z^0$ bosons where their masses are explained by the spin-0 Higgs ($H$) boson, and strong interactions are mediated by the exchange of gluons ($g$). Figure \ref{fig:SM} shows the elementary particle content of the SM. The first particle in Figure \ref{fig:SM} to be measured was the electron ($e$) in 1897 by J.~J.~Thomson and the last particle in Figure \ref{fig:SM} to be confirmed was the Higgs boson in 2012 by the ATLAS and CMS collaborations \cite{Higgs1:20121,Higgs2:201230}.

Leptons and quarks are spin-1/2 particles with six different types: electron ($e$), muon ($\mu$), tau ($\tau$), and the neutrinos $\nu_e, \ \nu_{\mu}, \ \nu_{\tau}$ are the six lepton types, and up $(u)$, charm $(c)$, top $(t)$, down $(d)$, strange $(s)$; and bottom $(b)$ are the six quark types (or \textit{flavours}). Quarks have fractional electric charge $q_e$, where $u$, $c$, and $t$ have $q_e=+\frac{2}{3}e$ and $d$, $s$, and $b$ have $q_e=-\frac{1}{3}e$, where $e$ is the elementary charge. 
Quark current masses can be found in Table \ref{tab:quark_masses}. Due to the large mass difference between quarks, one can divide them into light quarks ($u$, $d$, and $s$) and heavy quarks ($c$, $b$, and $t$). Furthermore, quarks (and gluons) are the only fundamental particles that participate in strong interactions. This thesis will focus on studying the strong interactions.

\begin{table}[ht]
	\centering
	\renewcommand{\arraystretch}{1.5}
	\begin{tabular}{cc} \hline
		Quark & Bare Mass (MeV) \\ \hline
		$u$ & 2.16\\
		$d$ & 4.67\\
		$s$ & 93.4\\
		$c$ & 1270\\
		$b$ & 4180\\
		$t$ & 173000\\ \hline
	\end{tabular}
	\caption[Current masses for the six flavours of quarks. The $u$, $d$, and $s$ masses are at 2 GeV renormalization scale; the $c$ and $b$ masses are defined by $\bar{m}=\bar{m}(\mu=\bar{m})$; and the $t$ mass is the pole mass. Masses are in natural units where $c=1$ and $\hbar=1$ (see Appendix \ref{Appendix_A} for the conventions used in this thesis).]{Current masses for the six flavours of quarks \cite{pdg:2022}. The $u$, $d$, and $s$ masses are at 2 GeV renormalization scale; the $c$ and $b$ masses are defined by $\bar{m}=\bar{m}(\mu=\bar{m})$; and the $t$ mass is the pole mass. Masses are in natural units where $c=1$ and $\hbar=1$ (see Appendix \ref{Appendix_A} for the conventions used in this thesis).}
	\label{tab:quark_masses}
\end{table}

\begin{figure}[ht]
	\centering
	\includegraphics{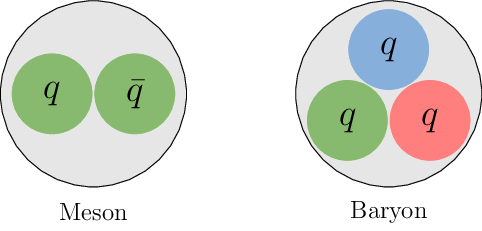}
	\caption{Hadronic content in the conventional quark model. Mesons are a $q\bar{q}$ bound state; baryons are a $qqq$ bound state. Colour combinations in the figure are defined in Section \ref{sec:quark_colour_group}.}
	\label{fig:SM_hadrons}
\end{figure}

Quarks were proposed as elementary particles by Gell-Mann \cite{Gell-Mann:1964ewy} and Zweig \cite{Zweig:1964jf} in 1964. In the quark model hadrons are bound states of quarks interacting via the strong interaction, and can be divided in mesons (integer spin) and baryons (half-odd integer spin). In the conventional quark model, the bound state of a quark-antiquark ($q\bar{q}$) pair is called a meson, for example a pion $\pi^+$ is a $u\bar{d}$ bound state. The bound state of three quarks ($qqq$) is called a baryon, for example protons are an $uud$ bound state and neutrons are an $udd$ bound state. Figure \ref{fig:SM_hadrons} shows the meson and baryon bound states. 

\begin{figure}[ht]
	\centering
\includegraphics{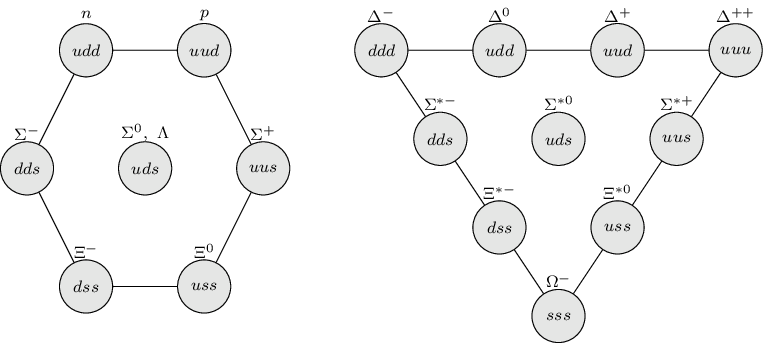}
	\caption{$J^P=\frac{1}{2}^+$ baryon octet (left) and $J^P=\frac{3}{2}^+$ baryon decuplet (right).}
	\label{fig:eightfoldway}
\end{figure}

The idea of hadrons as bound states of elementary particles explained the patterns observed in the eightfold way \cite{Gell-Mann:1961eightfold,Neeman:1961eightfold}. For instance, spin-1/2 and spin-3/2 baryons with $u$, $d$, or $s$ quarks can be arranged in an octet and in a decuplet, respectively, as shown in Figure \ref{fig:eightfoldway}. Due to the small light quark masses and flavour-independence of strong interactions, there is an approximate SU(3) flavour symmetry that allows us to form the multiplets structure observed in the eightfold way.

The apparent violation of the Pauli exclusion principle in the baryon $\Delta^{++}$ with three $u$ quarks with parallel spins is fixed when a new quantum number (colour) is added to quarks \cite{Greenberg:1964pe}. Quarks have colour charge red, blue, or green, and antiquarks have anti-colour charge anti-red, anti-blue, or anti-green. Gluons are also coloured and carry a colour and an anti-colour. The colour degrees of freedom are illustrated in Figure \ref{fig:SM_hadrons}.

\subsection{Colour Confinement and the SU(3) Colour Group}
\label{sec:quark_colour_group}
All conventional hadrons that have been observed are \textit{colourless}, meaning that they are formed by colour-anticolour (meson), three colours (baryon), or three anticolours (antibaryon) combinations. No free quark has ever been observed. This leads us to colour confinement, where since quarks have colour charge they are not observed free but instead confined in colourless hadrons. If one gives enough energy to a meson, for example, in order to split it in a quark and an antiquark, new quark-antiquark pairs are created, no free quarks are observed, and hadronic jets are observed in a process called hadronization \cite{narison}.

The strong interactions are described by a gauge theory with SU(3) colour group symmetry. In this group, quarks are in the fundamental representation \textbf{3}, antiquarks are in the conjugate representation $\mathbf{3^*}$, and gluons are in the adjoint representation \textbf{8}. Therefore, mesons are in the $\mathbf{3}\otimes\mathbf{3^*}$ and baryons are in the $\mathbf{3}\otimes\mathbf{3}\otimes\mathbf{3}$ product representations. These products are reducible and can be written in terms of irreducible representation as
\begin{align}
	\label{eq:irred_rep_conventional}
	\begin{split}
		&\mathbf{3}\otimes\mathbf{3^*} = \mathbf{1}\oplus\mathbf{8} \,, \\
		&\mathbf{3}\otimes\mathbf{3}\otimes\mathbf{3} = \mathbf{1}\oplus\mathbf{8}\oplus\mathbf{8}\oplus\mathbf{10} \,.
	\end{split}
\end{align}
The colour singlet \textbf{1} in Eq.~\eqref{eq:irred_rep_conventional} is what was previously called colourless hadron. Therefore, only colour singlet hadrons are experimentally observable, as represented in Figure \ref{fig:SM_hadrons}. Furthermore, Eq.~\eqref{eq:irred_rep_conventional} is the same group decomposition of the SU(3) flavour group, as can be seen, for example, in the flavour multiplet structures in Figure \ref{fig:eightfoldway}. 

\subsection{Angular Momenta, Parity, and Charge Conjugation}
Beyond the valence quark content and the electric charge, hadrons can be described as well by their total angular momenta, parity, and charge conjugation via the $J^{PC}$ quantum number. The total angular momentum is given by $\vec J = \vec L + \vec S$ where $\vec L$ is the orbital angular momentum and $\vec S$ is the spin angular momentum. Quarks are spin-1/2 particles, therefore mesons have spin quantum numbers $s = 0, 1$. The orbital quantum number is $l=0, 1, 2, \dots$, where $\vec L^2=l(l+1)$ and $\vec L=0$ corresponds to the ground state, and higher values of $\vec L$ correspond to excited states. States with $l=0, 1, \dots$ are also called S-waves, P-waves, and so on. Note that natural units are being used in this thesis (see Appendix \ref{Appendix_A} for conventions). In the conventional quark model, parity ($P$) and charge conjugation ($C$) are determined in the following way \footnote{Note that only flavour neutral mesons have definite charge conjugation $C$.}
\begin{align}
	\label{eq:jpc}
	\begin{split}
		P &= (-1)^{l+1} \\
		C &= (-1)^{l+s} \,.
	\end{split}
\end{align}
Therefore, there is a set of allowed $J^{PC}$ quantum numbers for mesons in the conventional quark model:
\begin{align}
	\label{eq:conventional_jpc}
	J^{PC} = 0^{-+},0^{++},1^{--},1^{+-},1^{++},2^{++}, \dots \,,
\end{align}
and any measurement of a meson with the $\textit{exotic}$ quantum numbers $J^{PC}=0^{--}$, $0^{+-}$, $1^{-+}$,  $2^{+-},\ldots$ is not predicted by the conventional quark model and it is an indication of a beyond the conventional quark model state.

In Chapter \ref{chapter:Paper_2}, the bound states studied are not eigenstates of the charge conjugation operator $C$, hence only $J$ and $P$ quantum numbers are defined. The quantum numbers found are the scalar $J^P = 0^+$, the pseudoscalar $J^P = 0^-$, the axial vector $J^P = 1^+$, and the vector $J^P = 1^-$.

\section{Beyond the Conventional Quark Model}
\label{Sec:BCQM_main_idea}
\begin{figure}[ht]
	\centering
	\includegraphics{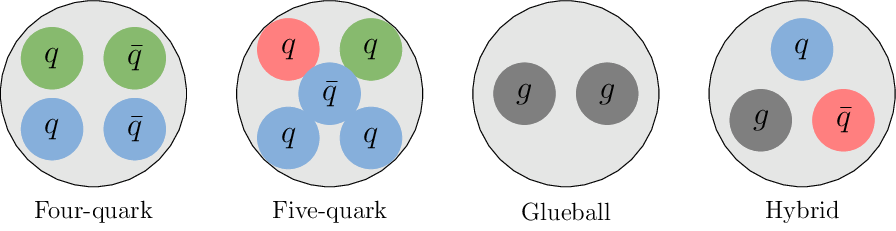}
	\caption{Beyond the conventional quark model hadronic content. Quarks $q$ have colour charge green, blue, or red; antiquarks $\bar q$ have anti-colour charge anti-red, anti-blue, or anti-green; coloured gluons are represented by a grey colour.}
	\label{fig:beyond_hadrons}
\end{figure}

In addition to mesons as $q\bar{q}$ states, baryons as $qqq$ and antibaryons as $\bar{q}\bar{q}\bar{q}$ combinations, colour singlet states can be formed by considering hadrons with more than two (mesons) or three (baryons) quarks and with valence gluons. The following products also produce colour singlets in the SU(3) colour group\footnote{Group theory results were obtained using LieArt \cite{Feger:2019tvk}; see, e.g., Ref.~\cite{Cheng:1984vwu} for a group theory review.}:
{\allowdisplaybreaks
	\begin{subequations}
		\begin{align}
			&\mathbf{3}\otimes \mathbf{3^*}\otimes \mathbf{3}\otimes \mathbf{3^*} = 2(\mathbf{1})\oplus 4(\mathbf{8})\oplus \mathbf{10}\oplus\mathbf{10^*} \oplus\mathbf{27} \,, \label{eq:irred_rep_beyond_conventional_fourquark}\\
			&\mathbf{3}\otimes\mathbf{3} \otimes\mathbf{3}\otimes \mathbf{3}\otimes\mathbf{3^*} = 3(\mathbf{1})\oplus 8(\mathbf{8})\oplus 4(\mathbf{10})\oplus 2(\mathbf{10^*})\oplus 3(\mathbf{27})\oplus\mathbf{35} \,, \label{eq:irred_rep_beyond_conventional_fivequark}\\
			&\mathbf{8}\otimes\mathbf{8} = \mathbf{1}\oplus 2(\mathbf{8})\oplus \mathbf{10}\oplus \mathbf{10^*}\oplus \mathbf{27} \,, \label{eq:irred_rep_beyond_conventional_glueball}\\
			&\mathbf{3}\otimes \mathbf{3^*}\otimes\mathbf{8} = \mathbf{1}\oplus 3(\mathbf{8})\oplus \mathbf{10}\oplus \mathbf{10^*} \oplus\mathbf{27} \,, \label{eq:irred_rep_beyond_conventional_hybrid}
		\end{align}    
\end{subequations}}%
where $2(\mathbf{1}) = \mathbf{1} \oplus\mathbf{1}$, $4(\mathbf{8}) = \mathbf{8} \oplus \mathbf{8} \oplus \mathbf{8} \oplus \mathbf{8}$, and so on. Eq.~\eqref{eq:irred_rep_beyond_conventional_fourquark} is a four-quark state $q\bar{q}q\bar{q}$, Eq.~\eqref{eq:irred_rep_beyond_conventional_fivequark} is a five-quark state $qqqq\bar{q}$, Eq.~\eqref{eq:irred_rep_beyond_conventional_glueball} is called a glueball $gg$, and Eq.~\eqref{eq:irred_rep_beyond_conventional_hybrid} is a hybrid state $q\bar{q}g$ where the quark-antiquark pair is not in a colour singlet state. Hadrons beyond the conventional quark model of $q\bar{q}$ mesons, $qqq$ baryons and $\bar{q}\bar{q}\bar{q}$ antibaryons are called \textit{exotic} hadrons. Figure \ref{fig:beyond_hadrons} shows some examples of exotic hadrons (see, e.g., Refs.~\cite{Olsen:newhadronspectroscopy2015,Olsen:RevModPhys.90.015003} for a review). It is important to note that exotic hadrons may or may not have exotic $J^{PC}$ quantum numbers.

The construction of those exotic hadrons were made considering only the colour confinement requirement in strong interactions. Therefore, the experimental detection of such states is expected if our understanding of colour confinement is correct. However, at the same time, the internal particle content of an exotic hadron can be difficult to determine as its internal structure becomes more complex. On the other hand, the measurement of a hadron with exotic quantum numbers could be a straightforward evidence of beyond the conventional quark model hadrons.

\subsection{Four-Quark States}
\begin{figure}[ht]
	\centering
	\includegraphics{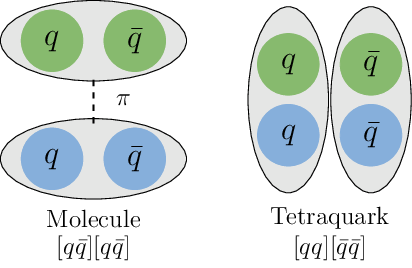}
	\caption{Tetraquark and molecular internal structure for four-quark states.}
	\label{fig:tetraquark}
\end{figure}

Four-quark states can be divided into two types based on their internal structure\footnote{Four-quark states can also be divided into four types if one consider their internal structure and interactions: tetraquark, molecules, adjoint-charmonium \cite{Braaten:2013boa}, and hadro-charmonium \cite{Dubynskiy:2008mq}.}. As shown in Figure \ref{fig:tetraquark}, there is a loosely bound molecular state $[q\bar{q}][q\bar{q}]$, where two mesons are bound at short-distances by a colour interaction and at large-distances by one-pion exchange \cite{Rujula:PhysRevLett.38.317,Tornqvist:ZPhysC61.525} and the tightly-bound tetraquark state $[qq][\bar{q}\bar{q}]$, composed by a coloured diquark $[qq]$ and a coloured antidiquark $[\bar{q}\bar{q}]$ bound through the colour force between the diquark and the antidiquark, and between each quark and each antiquark \cite{Anselmino:1992vg,Jaffe:2004ph,Santopinto:PhysRevC.75.045206}. Diquarks are the main objects that will be explored in this thesis. In Chapter \ref{chapter:Paper_2} the masses for diquarks with diverse valence quark content are calculated. Diquark masses provide QCD evidence/inputs for the constituent diquark models of tetraquark mesons (see, e.g., Ref.~\cite{Maiani:2004vq}).

\subsubsection{Diquarks}
A diquark (antidiquark) is a bound state of two quarks (antiquarks) \cite{Anselmino:1992vg,Jaffe:2004ph,Santopinto:PhysRevC.75.045206,JAFFE_2001}. In the SU(3) colour group it results in a antisymmetric triplet (antitriplet) $\mathbf{3^*}$ diquark, a symmetric sextet $\mathbf{6}$ diquark, a symmetric triplet $\mathbf{3}$ antidiquark, and a antisymmetric sextet (antisextet) $\mathbf{6^*}$ antidiquark:
\begin{align}
	\label{eq:irred_rep_beyond_conventional_diquark}
	\begin{split}
		&\mathbf{3}\otimes \mathbf{3} = \mathbf{3^*}\oplus \mathbf{6} \,, \\
		&\mathbf{3^*}\otimes \mathbf{3^*} = \mathbf{3}\oplus \mathbf{6^*} \,.
	\end{split} 
\end{align}
The same group decomposition of Eq.~\eqref{eq:irred_rep_beyond_conventional_diquark} can be done considering the SU(3) flavour group with the light quarks $u$, $d$, and $s$. 

\begin{figure}[ht]
	\centering
	\includegraphics{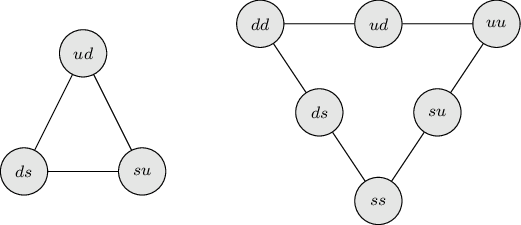}
	\caption{The $\mathbf{3^*}_f$ good (antisymmetric) and $\mathbf{6}_f$ bad (symmetric) diquarks quark content.}
	\label{fig:diquark_quark_content}
\end{figure}

Combining SU(3) colour and flavour groups the diquarks found are $(\mathbf{3^*}_c,\mathbf{3^*}_f)$, $(\mathbf{3^*}_c,\mathbf{6}_f)$, $(\mathbf{6}_c,\mathbf{3^*}_f)$, and $(\mathbf{6}_c,\mathbf{6}_f)$, where the subscripts $c$ and $f$ refer to colour and flavour, respectively. Diquarks in the $\mathbf{6}_c$ configuration are not favoured \cite{Jaffe:2004ph}, resulting in the $(\mathbf{3^*}_c,\mathbf{3^*}_f)$ and $(\mathbf{3^*}_c,\mathbf{6}_f)$ configurations. In Ref.~\cite{Jaffe:2004ph} $(\mathbf{3^*}_c,\mathbf{3^*}_f)$ is called a \textit{good} diquark while $(\mathbf{3^*}_c,\mathbf{6}_f)$ is called a \textit{bad} diquark due to the colour force between the quarks in the antisymmetric diquark being attractive \cite{JAFFE_2001}. Figure \ref{fig:diquark_quark_content} shows the quark content for the good and bad diquarks.

As seen in Eq.~\eqref{eq:irred_rep_beyond_conventional_diquark}, diquarks and antidiquarks are not colour singlet states and are, consequently, confined in hadrons where the $\mathbf{3^*}_c$ diquark, for example, can be combined with a $\mathbf{3}_c$ object in a similar way that a $\mathbf{3^*}$ antiquark is combined with a $\mathbf{3}$ object creating a colour singlet.

\subsection{\texorpdfstring{$XYZ$}{XYZ} States}
Mesons with a heavy quark-antiquark pair ($c\bar{c}$ or $b\bar{b}$) with properties that are not well described by the conventional quark model, such as mass, decay width, and electric charge are exotic mesons candidates and they are called $XYZ$ mesons\footnote{A new naming scheme for hadrons is being used by PDG \cite{pdg:2022} in order to accommodate exotic states.}. The internal structure and particle content of the $XYZ$ states is a topic of current interest (see, e.g., Refs.~\cite{Brambilla:2019esw,Swanson:2006st,Godfrey:2008nc, Olsen:2015zcy} for a review). They could be described for example, in addition to a conventional meson, by a hybrid $Q\bar{Q}g$, a $[\bar{q}Q][q\bar{Q}]$ molecular state, or a $[qQ][\bar{q}\bar{Q}]$ tetraquark, where $q$ represents the light quarks $u$, $d$ or $s$ and $Q$ represents the heavy quarks $c$ or $b$. 

The first $XYZ$ particle candidate to be measured was the $X(3872)$ in 2003 \cite{X3872:PhysRevLett.91.262001}. The observed decay processes $X(3872) \rightarrow \omega J/\psi$, $X(3872) \rightarrow \gamma J/\psi$ and $X(3872) \rightarrow \rho^0 J/\psi$ indicates that the $X(3872)$ have a $J^{PC} = 1^{++}$ \cite{belle:abe2005evidence,belle:abe2005experimental,LHCb:X3872PhysRevD.92.011102}. The first possibility for the quark content of this state is a $c\bar{c}$ pair, however some properties of the $X(3872)$ are in conflict with this interpretation \cite{Barnes:PhysRevD.69.054008,Swanson:2006st}. The decay processes also indicate that the state could be composed by $c\bar{c}u\bar{u}$ quarks, allowing a four-quark interpretation. The measured mass of the $X(3872)$ is close to the $D^0\bar{D}^{*0}$ threshold, which suggests that it might be a molecular state \cite{Tornqvist:x3872molecule,Swanson:x3872molecule}. However a tetraquark content also explain $X(3872)$ 
properties such as its parity \cite{Maiani:2004vq}.

An interesting example of a $XYZ$ state is the $Z^+$(4430) state. The $Z^+$(4430) state, measured by the Belle group in 2008 \cite{Belle:Z4430PhysRevLett.100.142001}, has nonzero electric charge and decays to $Z^+(4430)\rightarrow \pi^+ J/\psi$ \cite{Belle:Z4430PhysRevD.90.112009}, which is evidence that the $Z^+$(4430) particle must contain a $c\bar{c}$ pair. However a conventional $c\bar{c}$ meson and a $c\bar{c}g$ hybrid state are electrically neutral, which suggests that the $Z^+$(4430) state is a strong candidate for a four-quark meson.

\section{Quantum Chromodynamics}
\label{Sec:qcd_main_idea}

Quantum chromodynamics (QCD) is the theory for the strong interaction and describes quarks and gluons dynamics. QCD is a non-abelian SU(3) gauge theory where the gluon is the non-abelian gauge field mediating the colour interactions between quarks. The non-abelian property of the gluon field is a special feature resulting in the self-interaction of gluons, i.e., gluons carry colour charge, differently from photons in the electromagnetic interaction, for which the dynamics is described by an abelian gauge theory called quantum electrodynamics (QED). The gauge group of the unified electroweak interactions contains a SU(2)$_\text{L}\otimes$U(1)$_\text{Y}$ which holds both abelian and non-abelian symmetry.

In this chapter key points of QCD are introduced. These points are going to be important in the next chapters of this thesis when defining QCD sum-rules (QCDSR) and applying it to diquarks.
\subsection{QCD Lagrangian Density}
\label{Sec:qcd_lagrangian}
The colour force that bind quarks together, and consequently creates particles like the $XYZ$ mesons is described by the renormalizable non-abelian gauge theory quantum chromodynamics (QCD), where gluons are the particles responsible for the interactions. The quantized Lagrangian density for QCD is \cite{Pascual:1984zb}:
%{\allowdisplaybreaks
	{\allowdisplaybreaks\begin{align}
			\mathcal{L}_{QCD}(x)&= -\dfrac{1}{2}[\partial_{\mu}B^a_{\nu}(x)][\partial^{\mu}B^{\nu}_a(x)-\partial^{\nu}B^{\mu}_a(x)]-\dfrac{1}{2\xi}[\partial_{\mu}B^{\mu}_a(x)][\partial_{\nu}B^{\nu}_a(x)] \nonumber \\
			&+ \dfrac{i}{2} \bar{q}^A_{\alpha}(x)\gamma^{\mu}\partial_{\mu}q^A_{\alpha}(x)-\dfrac{i}{2} [\partial_{\mu}\bar{q}^A_{\alpha}(x)]\gamma^{\mu}q^A_{\alpha}(x) - m_A\bar{q}^A_{\alpha}(x)q^A_{\alpha}(x) \nonumber \\
			&+ \dfrac{1}{2} g \bar{q}^A_{\alpha}(x) \lambda^a_{\alpha\beta}\gamma^{\mu}q^A_{\beta}(x)B^a_{\mu}(x) \nonumber \\
			&- \dfrac{1}{2}gf_{abc}[\partial_{\mu}B^a_{\nu}(x)-\partial_{\nu}B^a_{\mu}(x)]B^{\mu}_b(x)B^{\nu}_c(x) \nonumber \\
			&-\dfrac{1}{4}g^2f_{abc}f_{ade}B^b_{\mu}(x)B^c_{\nu}(x)B^{\mu}_d(x)B^{\nu}_e(x) \nonumber \\
			&-[\partial_{\mu}\bar{\phi}_a(x)]\partial^{\mu}\phi_a(x)+gf_{abc}[\partial_{\mu}\bar{\phi}_a(x)]\phi_b(x)B^{\mu}_c(x) \,,
			\label{eq:qcd_lagrangian}
	\end{align}}%
	where $q^A_{\alpha}(x)$ are Dirac spinor fields representing the quarks, $B^{\mu}_a(x)$ are gauge fields representing the gluons and $\phi_a(x)$ are the non-physical Faddeev-Popov ghosts introduced during the quantization of the QCD Lagrangian density. The indices $\mu$, and $\nu$ are space-time indices. The index $A=1,2,\ldots,N_f$ is a quark flavour index, where $N_f$ is the number of flavours. The index $\alpha=1,2,\ldots,N_c$ is the quark colour index, where $N_c$ is the number of quark colours, and the index $a=1,2,\ldots N_c^2-1$ is the gluon colour index. Furthermore, $\gamma^{\mu}$ are the Dirac matrices, $m_A$ is the quark mass, $g$ is the strong coupling, $\xi$ is a gauge parameter and $\lambda_a = 2 t_a$, where $t_a$ are the colour group $SU(N_c)$ generators and satisfy $[t_a,t_b]=if_{abc}t_c$. See Appendix \ref{Appendix_A} for a more detailed explanation on the notation and conventions used in this thesis.
	
	The first two lines in the Lagrangian density (\ref{eq:qcd_lagrangian}) are interpreted as the gluon and quark field propagators, respectively. The third line represents the interaction among quarks and gluon fields. The fourth and fifth lines shows the interaction terms between three gluon fields and four gluon fields, respectively, due to the non-abelian feature of QCD. The sixth line represents the ghost propagator and interaction with gluon fields.
\subsection{Two-point Correlation Function}
\label{sec:two-point_correlation_function}

The two-point correlation function gives the propagation amplitude of a particle between two space-time points \cite{peskin} and it is written as the vacuum expectation value of fields
\begin{align}
	\langle\Omega|Tq(x)\bar{q}(y)|\Omega\rangle \,,
	\label{eq:correlator}
\end{align}
where $|\Omega\rangle$ is the vacuum of the interacting theory, $T$ is the time-ordering operator, and $q$ represents a fermionic field (a quark field, for example), where possible flavour and/or colour indices are omitted.
For a free (non-interacting) theory $|\Omega\rangle$ is replaced by $|0\rangle$, the free-theory vacuum, and Eq.~\eqref{eq:correlator} can be written as
\begin{align}
	\langle0|Tq(x)\bar{q}(y)|0\rangle\,.
	\label{eq:free_correlator}
\end{align}
According to Wick's theorem \cite{peskin},
\begin{align}
	Tq(x)\bar{q}(y)=N\left[q(x)\bar{q}(y)+\text{all possible contractions}\right]
	\label{eq:Wicks_theorem}
\end{align}
where $N$ is the normal-ordering operator. Another notation for the normal-ordering operator found in the literature and in this thesis is $\langle0|N[AB]|0\rangle\equiv\  \langle0|:AB:|0\rangle$. The free-theory vacuum expectation value of any uncontracted field or product of fields is zero at normal order, therefore Eq.~\eqref{eq:free_correlator} gives
\begin{align}
	\begin{split}
		\langle0|Tq(x)\bar{q}(y)|0\rangle &\equiv \contraction[1ex]{}{q}{(x)}{\bar{q}}q(x)\bar{q}(y) = i S^{(0)}(x-y) \\
		&= \frac{i}{(2\pi)^4}\int\, d^4p \ \frac{\slashed{p}+m}{p^2-m^2+i\epsilon}\, e^{-i p\cdot (x-y)} \,,    
	\end{split}
	\label{eq:free_propagator}
\end{align}
where $S^{(0)}(x-y)$ is the free-field propagator, $p$ is an internal momentum, $\slashed{p}=\gamma^{\mu}p_{\mu}$ is the Feynman slash notation, $\epsilon\rightarrow0^+$, and the last equality is obtained calculating the Fourier transform of the propagator $S^{(0)}(x-y)$. 

It can be shown \cite{peskin,Pascual:1984zb} that the interacting theory propagator, Eq.~(\ref{eq:correlator}), can be related to Eq.~(\ref{eq:free_correlator}) via
\begin{align}
	\label{eq:gellmann_low_relation}
	\langle\Omega|Tq(x)\bar{q}(y)|\Omega\rangle = \frac{\langle0|Tq(x)\bar{q}(y) \exp[i\int d^dz\mathcal{L}_{\text{int}}(z)]|0\rangle}{\langle0|T\exp[i\int d^dz\mathcal{L}_{\text{int}}(z)]|0\rangle} \,,
\end{align}
where $\mathcal{L}_{\text{int}}(z)$ is the interaction part of the Lagrangian density. The numerator of Eq.~\eqref{eq:gellmann_low_relation} represents the sum of all connected two-point Feynman diagrams times the exponential of the sum of all two-point disconnected Feynman diagrams. The denominator of Eq.~\eqref{eq:gellmann_low_relation} represents the exponential of the sum of all two-point disconnected Feynman diagrams. Therefore, the contributions from disconnected diagrams cancel, and Eq.~\eqref{eq:gellmann_low_relation} represents the sum of all connected one-particle irreducible (1PI) two-point Feynman diagrams (diagrams that cannot be split in two disconnected diagrams by removing an internal line).

Considering only connected Feynman diagrams and using perturbation theory, the exponential $e^{i\int d^dz\mathcal{L}_{\text{int}}(z)}$ in Eq.~\eqref{eq:gellmann_low_relation} can be expanded as a power series as
\begin{align}
	\begin{split}
		\label{eq:correlator_expansion}
		\langle\Omega|Tq(x)\bar{q}(y)|\Omega\rangle &= \langle0|Tq(x)\bar{q}(y)|0\rangle \\
		&+ \langle0|Tq(x)\bar{q}(y)\left[i\int d^dz\mathcal{L}_{\text{int}}(z)\right]|0\rangle \\
		&+ \langle0|Tq(x)\bar{q}(y)\left[\frac{i^2}{2!}\int d^dz\ d^dw\mathcal{L}_{\text{int}}(z)\mathcal{L}_{\text{int}}(w)\right]|0\rangle + \cdots \,.
	\end{split}
\end{align}
Therefore, the two-point correlation function in the interacting theory, Eq.~\eqref{eq:correlator}, has as its first term Eq.~\eqref{eq:free_propagator}, the two-point correlation function for the free theory, followed by terms that are corrections to the free propagator. These corrections are represented by Feynman diagrams with increasing loop order as the order of the expansion in Eq.~\eqref{eq:correlator_expansion} increases (e.g., a two-loop diagram involves integration over two independent four-momenta variables). Note that $\mathcal{L}_{\text{int}}$ is proportional to the strong coupling $g$ for the Lagrangian density \eqref{eq:qcd_lagrangian} consequently, increasing the order of the expansion \eqref{eq:correlator_expansion} means to increase the order in $g$. In Section \ref{Sec:qcd_asymp_freedom} it is going to be shown that asymptotic freedom is what guarantees that perturbation theory is valid for QCD in the high-energy range.
\subsection{Composite Operators}
\label{Sec:composite_operators}

In the QCD sum-rules analysis we are interested in calculating the correlation function of composite operators,
\begin{align}
	\langle\Omega|TJ(x)J^{\dagger}(0)|\Omega\rangle \,,
	\label{eq:correlator_composite}
\end{align}
where the current $J(x)$ is a composite operator, which means that $J(x)$ is the product of fields at the same space-time point $x$. The currents $J(x)$ and $J(0)$ are chosen to have the same quantum numbers $J^{PC}$ (by writing the appropriate Lorentz structure) and valence quark content as the system being studied. The currents $J^{\dagger}(0)$ and $J(x)$ can be interpreted as operators that create the states with the same quantum numbers as the current at the space-time point $0$, and then annihilate the state at the space-time point $x$.

For instance, in Chapter \ref{chapter:Paper_2}, we have
\begin{align}
	\label{eq:heavy_light_current}
	J_{\alpha}(x) = \epsilon_{\alpha\beta\gamma}Q^{\text{T}}_{\beta}(x)C q_{\gamma}(x)
\end{align}
as a current that describes a heavy-light pseudoscalar ($J^P=0^-$) diquark, where $\epsilon_{\alpha\beta\gamma}$ is the Levi-Civita symbol with quark colour indices $\alpha$, $\beta$, and $\gamma$, $Q_{\beta}$ is a heavy-quark field (charm $c$ or bottom $b$), $T$ is the transpose, $C$ is the charge conjugation operator, and $q_{\gamma}$ is a light-quark field.

The correlation function \eqref{eq:correlator_composite} can be expanded as a power series as in Section \ref{sec:two-point_correlation_function}, and it is obtained in the momentum space by calculating its Fourier transform: 
\begin{align}
	\Pi(Q^2)=i\int d^dx e^{iq\cdot x}\langle\Omega|TJ(x)J^{\dagger}(0)|\Omega\rangle \,,
	\label{eq:correlator_composite_momentum}
\end{align}
where $Q^2=-q^2$ is the Euclidean external momentum. The correlation function written in the form of Eq.~\eqref{eq:correlator_composite_momentum} is one of the fundamental pieces of QCDSR in Section \ref{sec:qcdsr_main_idea} and it carries all the QCD information needed about the system being studied.
\subsection{Regularization and Renormalization}
\label{Sec:qcd_regularization}
After the Fourier transform of Eqs.~\eqref{eq:correlator_expansion} or \eqref{eq:correlator_composite}, the two-point correlation function $\Pi(Q^2)$ is an integration over the internal momenta that might be a divergent integral when the space-time dimension $d$ is equal to 4. To be able to control such divergences a regularization method need to be used. One example of regularization method is dimensional regularization \cite{tHooft:1972tcz} (see Ref.~\cite{peskin} for a review) where the space-time dimension is kept to be $d$ during the integral evaluation. After that, the limit $d\rightarrow4+2\epsilon$ is taken and the two-point correlation function is expanded around $\epsilon=0$ resulting in an expression where the divergences appear as singularities as $\epsilon\rightarrow0$. In QCD the most common regularization method is the dimensional regularization because it preserves gauge invariance \cite{Pascual:1984zb}.

However, experimentally measured quantities, such as masses and coupling constants, are finite. To eliminate the regularized divergences a renormalization method need to be applied to the theory. In the conventional renormalization method counterterms $C_i$ are added to the Lagrangian density \eqref{eq:qcd_lagrangian} as renormalization constants $Z_i$, where $Z_i\equiv 1-C_i$, creating new contributions to the theory. These counterterms are then chosen in such a way that the regularized divergences are eliminated from the theory. 

Different ways of choosing $C_i$ are called different renormalization schemes. In QCD the minimal subtraction renormalization scheme (MS-scheme) and the modified minimal subtraction renormalization scheme ($\overline{\text{MS}}$-scheme) are widely used. In the former all counterterms $C_i$ are a power series expansion in the dimensionless coupling $\alpha$, where
\begin{align}
	\alpha\equiv \frac{(g\mu^{\epsilon})^2}{4\pi}
\end{align}
and $\mu$ is a renormalization mass scale with mass dimension. In the latter the $-\ln{4\pi}+\gamma_E$ constant that appears in the calculation is eliminated from the renormalized theory by making the substitution $1/\epsilon\rightarrow 1/\epsilon-\ln{4\pi}+\gamma_E$ (or equivalently, $\mu^2\rightarrow\frac{e^{\gamma_E}}{4\pi}\mu^2$), where $\gamma_E$ is the Euler-Mascheroni constant. Note that, with the space-time dimension equals to $d$, the coupling $g$ has dimension $[\text{M}]^{\frac{4-d}{2}}$ (see Appendix \ref{Appendix_A}) and, consequently, scale invariance of $g$ is broken when $d\neq4$. The $\overline{\text{MS}}$-scheme is the renormalization scheme that is going to be used in this thesis. A review on regularization and renormalization can be found in Refs.~\cite{peskin,Pascual:1984zb,narison}.

The conventional renormalization process is more complicated when applied to composite operators due to the mixing of operators \cite{Narison:1983kn}. In Chapter \ref{chapter:Paper_1} the diagrammatic renormalization method is developed for the renormalization of composite operators. In this method each Feynman diagram is renormalized instead of adding counterterms to the theory Lagrangian density. It is shown that with the diagrammatic renormalization method the mixing of composite operators is automatically avoided, increasing, for example, the computational efficiency of the renormalization process. The diagrammatic renormalization method is the method used in Chapter \ref{chapter:Paper_2} for the renormalization of diquarks two-point correlation function within QCDSR.
\subsection{Asymptotic Freedom}
\label{Sec:qcd_asymp_freedom}
The strength of an interaction is characterized by its coupling. In the QCD Lagrangian density $g$ is the only coupling in all interaction terms of Eq.~\eqref{eq:qcd_lagrangian}. The process of regularization and renormalization, however, introduce the scale $\mu$ in the quantum field theory and, consequently, the coupling is modified by a scale dependency and it is called running coupling.

This scale dependence leads to interesting consequences for the interactions. For example, if the theory has a running coupling that tends to zero as the energy scale increases, the interaction strength likewise tends to zero and the theory is asymptotically free. On the other hand, if the running coupling increases as the energy scale increases, the theory has a Landau pole and it is well defined only up to some intermediate scale. A third option is that the running coupling reaches, at some intermediate scale $\Lambda$, a fixed value which is neither zero nor infinity; for any energy scale higher than $\Lambda$ the running coupling is scale-invariant and its value is equal to the value at the intermediate scale $\Lambda$, $\bar{g}(\Lambda)$. When this happens the theory is said to be asymptotically safe.

In Section \ref{Renormalization Group Equations} the beta function (the variation of the running coupling with respect to the energy scale) is analysed when it is positive, negative, and zero. A zero beta function means that the running coupling is scale invariant and has a fixed value called a fixed point.

\subsubsection{Renormalization Group Equation}
\label{Renormalization Group Equations}
Consider the renormalized Green function $G(p_i, \alpha, a, m, \mu)$ where $p_i = p_1, p_2, \dots, p_N$ are $N$ external momenta, $\alpha$ is the dimensionless coupling constant, $a$ is the gauge parameter, $m$ is a mass, and $\mu$ is a renormalization scale ($\mu$ is a renormalization mass scale in the MS- and $\overline{\text{MS}}$-scheme). The relation between the renormalized and the bare Green function is given by
\begin{align}
	G(p_i, \alpha, a, m, \mu) = Z(\mu) \ G_0(p_i, \alpha_0, a_0, m_0) \,,
	\label{renGreenFunction}
\end{align}
where $Z(\mu)$ is the renormalization constant determined by the field content of the Green function, and $\alpha_0$, $a_0$, and $m_0$ are the bare quantities for $\alpha$, $a$, and $m$, respectively. Note that the bare Green function $G_0(p_i, \alpha_0, a_0, m_0)$ is independent of the renormalization scale $\mu$, therefore, differentiating $G_0(p_i, \alpha_0, a_0, m_0)$ with respect to $\mu$ one has
\begin{align}
	\mu\frac{d}{d\mu} G_0(p_i, \alpha_0, a_0, m_0) =0 \,,
	\label{differentialBareG}
\end{align}
where, for convention, both sides of Eq.~\eqref{differentialBareG} were multiplied by $\mu$. Differentiating Eq.~\eqref{renGreenFunction} with respect to $\mu$ one finds
\begin{align}
	\begin{split}
		\mu\frac{d}{d\mu}G(p_i, \alpha, a, m, \mu)&=\mu\frac{d}{d\mu}\left[Z(\mu) \ G_0(p_i, \alpha_0, a_0, m_0)\right] \\
		&=\mu\frac{d}{d\mu}\left[Z(\mu)\right] G_0(p_i, \alpha_0, a_0, m_0) \\ 
		&= \mu\frac{d}{d\mu}\left[Z(\mu)\right] \frac{1}{Z(\mu)}\ G(p_i, \alpha, a, m, \mu)\,,
	\end{split}
	\label{differentialRenormalizedG}
\end{align}
where both sides of Eq.~\eqref{differentialRenormalizedG} were multiplied by $\mu$, Eq.~\eqref{differentialBareG} was used on the second line, and Eq.~\eqref{renGreenFunction} was used on the third line. Applying the chain rule in Eq.~\eqref{differentialRenormalizedG} one has
\begin{align}
	\mu\left[\frac{\partial}{\partial\mu}+\frac{d\alpha}{d\mu}\frac{\partial}{\partial \alpha}+\frac{1}{m}\frac{d m}{d\mu}m\frac{\partial}{\partial m} + \frac{d a}{d\mu}\frac{\partial}{\partial a}-\frac{1}{Z(\mu)}\frac{d Z(\mu)}{d\mu}\right]G(p_i, \alpha, a, m, \mu)=0\,.
	\label{DifferentialEquationRenorG}
\end{align}
From Eq.~\eqref{DifferentialEquationRenorG} one can define the functions
\begin{align}
	\begin{split}
		\alpha\beta(\alpha, a, x) &\equiv \mu\frac{d\alpha}{d\mu} \,, \\
		a\delta(\alpha, a, x) &\equiv \mu\frac{d a}{d\mu} \,, \\
		\gamma_m(\alpha, a, x) &\equiv -\frac{\mu}{m}\frac{d m}{d\mu} \,, \\
		\gamma_z(\alpha, a, x) &\equiv \frac{\mu}{Z(\mu)}\frac{d Z(\mu)}{d\mu} \,,
	\end{split}
	\label{UniversalFunctions}
\end{align}
where $x=m/\mu$, and rewrite Eq~\eqref{DifferentialEquationRenorG} as
{\allowdisplaybreaks
	\begin{align}
		\begin{split}  
			\biggl[\mu\frac{\partial}{\partial\mu}+\alpha\beta(\alpha, a, x)\frac{\partial}{\partial \alpha}&-\gamma_m(\alpha, a, x)x\frac{\partial}{\partial x} + a\delta(\alpha, a, x)\frac{\partial}{\partial a} \\
			&- \gamma_z(\alpha, a, x)\biggr]G(p_i, \alpha, a, m, \mu)=0\,.
		\end{split}
		\label{DifferentialEquationRenorGUniversalFunctions}
\end{align}}%
In the MS- and $\overline{\text{MS}}$-scheme Eq.~\eqref{UniversalFunctions} is independent of $m/\mu$ and $\beta(\alpha, a, m/\mu)$ is independent of the gauge parameter $a$ \cite{Pascual:1984zb,muta}. Therefore, one can write Eq.~\eqref{DifferentialEquationRenorGUniversalFunctions} as
{\allowdisplaybreaks
	\begin{align}
		\begin{split}
			\biggl[\mu\frac{\partial}{\partial\mu}+\alpha\beta(\alpha)\frac{\partial}{\partial \alpha}-\gamma_m(\alpha, a)x\frac{\partial}{\partial x} + a\delta(\alpha, a)\frac{\partial}{\partial a} -\gamma_z(\alpha, a)\biggr]G(p_i, \alpha, a, m, \mu)=0\,,    
		\end{split}
		\label{DifferentialEquationRenorGUniversalFunctionsAlpha}
\end{align}}%
where $\beta$ is called the beta function and $\gamma_z$ is called the anomalous dimension. Writing the dimension of $G(p_i)$ as $\left[G(p_i)\right] \equiv M^{d_G}$, one has
\begin{align}
	G(p_i, \alpha, a, m, \mu) \equiv \mu^{d_G}\ \overline{G}(p_i/\mu, \alpha, a, x)\,,
	\label{DimensionlessG}
\end{align}
where $d_G$ is the mass dimension of $G(p_i)$ and $\overline{G}$ is a dimensionless function. Introducing a dimensionless scale parameter $\lambda$ to Eq.~\eqref{DimensionlessG} one finds
{\allowdisplaybreaks\begin{align}
		\begin{split}
			G(\lambda p_i, \alpha, a, m, \mu) &= \mu^{d_G}\ \overline{G}(\lambda p_i/\mu, \alpha, a, x)\\
			&= \lambda^{d_G}\left(\frac{\mu}{\lambda}\right)^{d_G}\ \overline{G}(\lambda p_i/\mu, \alpha, a, x)\\
			&= \lambda^{d_G}\ \overline{G}(p_i, \alpha, a, m/\lambda, \mu/\lambda)\,.
		\end{split}
		\label{ScaleG}
\end{align}}
Differentiating Eq.~\eqref{ScaleG} with respect to $\lambda$ and multiplying both sides by $\lambda$, one obtains
{\allowdisplaybreaks\begin{align}
		\begin{split}
			\lambda \frac{\partial}{\partial\lambda}G(\lambda p_i, \alpha, a, m, \mu) &= \lambda \frac{\partial}{\partial\lambda}\left[\lambda^{d_G}\ \overline{G}(p_i, \alpha, a, m/\lambda, \mu/\lambda)\right] \\
			&= \left[\lambda\ d_G\ \lambda^{d_G-1} + \lambda^{d_G+1}\frac{\partial\left(m/\lambda\right)}{\partial\lambda}\frac{\partial}{\partial \left(m/\lambda\right)}\right. \\
			&+ \left.\lambda^{d_G+1}\frac{\partial\left(\mu/\lambda\right)}{\partial\lambda}\frac{\partial}{\partial \left(\mu/\lambda\right)}\right]\overline{G}(p_i, \alpha, a, m/\lambda, \mu/\lambda) \\
			&= \lambda^{d_G}\left[d_G - x\frac{\partial}{\partial x} - \mu\frac{\partial}{\partial \mu}\right]\overline{G}(p_i, \alpha, a, m/\lambda, \mu/\lambda) \\
			&= \left[d_G - x\frac{\partial}{\partial x} - \mu\frac{\partial}{\partial \mu}\right]G(\lambda p_i, \alpha, a, m, \mu)\,,
		\end{split}
		\label{ScaleGDifferentiation}
\end{align}}
where Eq.~\eqref{DimensionlessG} was used on the fifth line. Eq.~\eqref{ScaleGDifferentiation} can also be written as
\begin{align}
	\left[\lambda \frac{\partial}{\partial\lambda}-d_G + x\frac{\partial}{\partial x} + \mu\frac{\partial}{\partial \mu}\right]G(\lambda p_i, \alpha, a, m, \mu)=0\,.
	\label{ScaleGDifferentialForm}
\end{align}
In the MS and $\overline{\text{MS}}$ schemes, one can equate Eqs.~\eqref{DifferentialEquationRenorGUniversalFunctionsAlpha} and \eqref{ScaleGDifferentialForm} to obtain
\begin{align}
	\begin{split}
		\biggl[-\lambda \frac{\partial}{\partial\lambda}+\alpha\beta(\alpha)\frac{\partial}{\partial \alpha}+ a\delta(\alpha, a)\frac{\partial}{\partial a} &- \left\{1+\gamma_m(\alpha, a)\right\}x\frac{\partial}{\partial x}\\
		&+d_G-\gamma_z(\alpha, a)\biggr]G(\lambda p_i, \alpha, a, m, \mu)=0\,.
	\end{split}
	\label{ScaleGBoth}
\end{align}

Defining a parameter $t$ as $t \equiv \ln{\lambda}$, and the running coupling $\bar{\alpha}(t)$, the running gauge parameter $\bar{a}(t)$, and the running mass $\bar{x}(t)$ as the renormalized coupling constant, gauge parameter, and mass, respectively, Eq.~\eqref{UniversalFunctions} can be written as
\begin{align}
	\begin{split}
		\bar{\alpha}(t)\beta[\bar{\alpha}(t), \bar{a}(t), \bar{x}(t)] &= \frac{d\bar{\alpha}(t)}{dt} \,, \\
		\bar{x}(t)\gamma_m[\bar{\alpha}(t), \bar{a}(t), \bar{x}(t)] &= -\bar{x}(t) -\frac{d \bar{x}(t)}{dt} \,, \\
		\bar{a}(t)\delta[\bar{\alpha}(t), \bar{a}(t), \bar{x}(t)] &= \frac{d\bar{a}(t)}{dt} \,,
	\end{split}
	\label{RunningFunctions}
\end{align}
where $\bar{\alpha}(0)=\alpha$, $\bar{a}(0)=a$, and $\bar{x}(0)=x$.

%Using the running parameters, Eq.~\eqref{ScaleGBoth} is
%\begin{align}
%\begin{split}
%     \biggl[-\frac{\partial}{\partial t} +\beta(\bar{\alpha}(t))\frac{\partial}{\partial\bar{g}(t)} &+ 
%     \delta(\bar{g}(t), \bar{a}(t))\frac{\partial}{\partial \bar{a}(t)} \\
%     &- \left\{1+\gamma_m(\bar{g}(t), \bar{a}(t))\right\}\bar{m}(t)\frac{\partial}{\partial \bar{m}(t)}\\
%     &+d_G-\gamma_z(\bar{g}(t), \bar{a}(t))\biggr]G(e^t p_i, \bar{g}(t), \bar{a}(t), \bar{m}(t), \mu)=0\,.   
%\end{split}
%\label{RunningDifferentialEquation}
%\end{align}
The behaviour of a quantum field theory at small distances can be investigated by solving Eq.~\eqref{RunningFunctions} for the running coupling \cite{Gellmann:PhysRev.95.1300, Callan:PhysRevD.2.1541, Symanzik:CommMathPhys18.227}. Note that, as mentioned before, in the MS and $\overline{\text{MS}}$ schemes, the beta function is independent of the gauge parameter and the mass. Therefore, the differential equation for the running coupling can be written as
\begin{align}
	\beta[\bar{g}(t)] &= \frac{d\bar{g}(t)}{dt}\,.
\end{align}
in terms of the coupling $g$.

If $\beta[\bar{g}(t)]$ is greater than zero, the variation of the running coupling with respect to $t$ is greater than zero, therefore $\bar{g}(t)$ increases as $t\rightarrow\infty$. On the other hand, if $\beta[\bar{g}(t)]$ is negative, $\bar{g}(t)$ increases as $t\rightarrow-\infty$. If $\beta[\bar{g}(t)]$ is zero at $\bar{g}(t)=g^*$, the running coupling is scale invariant, and the points $g^*$ are called fixed points \cite{Wilson:PhysRevD.3.1818}. A fixed point can be a trivial (non-interacting) fixed point, where $\beta(g^*)=0$ when $g^*=0$, or a nontrivial (interacting) fixed point, where $\beta(g^*)=0$ when $g^*\neq0$. 

Consider the case where the beta function is positive and $g^*=0$ is a fixed point. In this case, $g^*$ is called an infrared (IR) fixed point because, as $t\rightarrow -\infty$ (the IR limit), the running coupling $\bar{g}(t)\rightarrow g^*=0$. As $t$ goes to the ultraviolet (UV) region, on the other hand, it can lead us to a Landau pole, where $\bar{g}(t)\rightarrow \infty$. This pattern for the running coupling can be found in QED \cite{Gellmann:PhysRev.95.1300}.

When the beta function is negative and $g^*=0$ is a fixed point as $t\rightarrow\infty$ (the UV limit), the running coupling goes to zero and $g^*$ is called an UV fixed point. A theory with this behaviour is an asymptotically free theory \cite{Gross:PhysRevLett.30.1343,Politzer:PhysRevLett.30.1346}. Furthermore, as $t$ goes to the IR region, $\bar{g}(t)\rightarrow \infty$. 

For a non-abelian gauge theory such as QCD, at the lowest order in perturbation theory, the running coupling is given by \cite{Pascual:1984zb}
\begin{align}
	\bar{\alpha}(t) &= \frac{\alpha}{1-\frac{\alpha}{\pi}\beta_1t} \,, 
\end{align}
where $\bar{\alpha}(0)=\alpha$, and
\begin{align}
	\beta_1 &= -\frac{1}{2}\left(11\frac{N_c}{3}-2\frac{N_f}{3}\right) \,,
\end{align}
with $\beta(\alpha)=\frac{\alpha}{\pi}\beta_1+\mathcal{O}(\alpha^2)$. For QCD, where $N_c=3$ and $N_f=6$, the beta function is negative with trivial fixed point; therefore the theory exhibits asymptotic freedom in the UV region \cite{Gross:PhysRevLett.30.1343,Politzer:PhysRevLett.30.1346}. It is because QCD is asymptotically free that perturbation theory can be used in high-energy scales. In Section \ref{sec:qcdsr_main_idea} we will see how QCD sum-rules approaches the region of low energy scales where non-perturbative effects become important.

\section{QCD Sum-Rules}
\label{sec:qcdsr_main_idea}
In the QCD sum-rule method the renormalized two-point correlation function of composite operators is connected to a hadronic spectral function through a dispersion relation allowing the extraction of hadronic properties \cite{Shifman:1978bx,Shifman:1978by} (see Refs.~\cite{Reinders:1984sr,Pascual:1984zb,deRafael:1997ea,Colangelo:2000dp,narison,Gubler:2018ctz} for a review). Due to the non-perturbative behaviour of the QCD coupling at low energies, the two-point correlation function of composite operators is calculated using the operator product expansion (OPE), where non-perturbative effects are incorporated in the theory by nonzero vaccum expectation values (VEVs) of local operators. Subtraction constants (including possible divergences) in the dispersion relation are eliminated by taking a sufficient number of $Q^2$ derivatives or by applying an integral transform.

\subsection{The Dispersion Relation}
\label{Sec:qcdsr_dispersion_relations}
QCD sum-rules are built considering quark-hadron duality where, via a dispersion relation, 
\begin{align}
	\label{eq:qcdsr_disp_rel}
	\begin{split}
		\Pi(Q^2) &= \Pi(0) + Q^2\Pi'(0) + \frac{1}{2}Q^4\Pi''(0) + \cdots \\
		&+ \frac{1}{n}Q^{2n}\Pi^{(n)}(0) + Q^{2n+2} \int_{t_0}^{\infty} \frac{\rho^{\text{had}}(t)}{t^{n+1}(t+Q^2)}dt \,,
	\end{split}
\end{align}
the renormalized two-point correlation function of composite operators ($\Pi(Q^2)$ on LHS) is related to a hadronic spectral function ($\rho^{\text{had}}(t)$ on RHS). In other words, an expression calculated using QCD (quarks and gluons) can be used to access information about hadrons through the dispersion relation \eqref{eq:qcdsr_disp_rel}. In Eq.~\eqref{eq:qcdsr_disp_rel} $t_0$ is the hadronic threshold, $\rho^{\text{had}}(t)$ is the hadronic spectral function, and $\Pi^{(n)}(0) = \left(\frac{d}{dQ^2}\right)^n \Pi(Q^2)\Bigr|_{Q^2=0}$ are subtraction constants that are usually unknown and divergent. 

The QCD side (LHS of Eq.~\eqref{eq:qcdsr_disp_rel}) is calculated using the OPE, and the hadronic side (RHS of Eq.~\eqref{eq:qcdsr_disp_rel}) is calculated using an experimentally known quantity or a model for the hadronic spectral function $\rho^{\text{had}}(t)$, as we shall see on Secs.~\ref{Sec:qcdsr_ope} and \ref{Sec:qcdsr_hadronic_spectral_function}. Furthermore, the possible divergent subtraction constants $\Pi^{(n)}(0)$ are local divergences (polynomials in $Q^2$) and can be eliminated via a sufficient number of $Q^2$ derivatives.
\subsection{Operator Product Expansion}
\label{Sec:qcdsr_ope}

The power series expansion in Eq.~\eqref{eq:correlator_expansion} is valid when $Q^2=-q^2$ is large, where $Q^2$ is the Euclidian momentum, due to the behaviour of the strong coupling $\alpha_s$. In the infrared region, where bound states of quarks and gluons are formed and $\alpha_s$ becomes large, non-perturbative effects need to be addressed. This can be done by considering that the vacuum expectation value of uncontracted fields in the correlation function is nonzero. The correlation function is then calculated using the operator product expansion (OPE) \cite{Wilson:PhysRev.179.1499}:
\begin{align}
	\label{eq:ope}
	\langle\Omega |:\mathcal{O}(x) \mathcal{O}(y) :|\Omega\rangle = \lim_{x\rightarrow y} \sum_n C_n(x-y) \langle\Omega|:\mathcal{O}_n(y):|\Omega\rangle\,,
\end{align}
where the local operators $\mathcal{O}$ are fields or products of fields, $C_n$ are constants called Wilson coefficients, and $: \ :$ represents the normal-ordering operator. The OPE expands the non-local vacuum expectation value of the operators $\mathcal{O}$ as a sum of local vacuum expectation values, ${\langle\Omega|:\mathcal{O}_n(y):|\Omega\rangle}$, called QCD \textit{condensates}. As the mass dimension $n$ of the local operators $\mathcal{O}_n(y)$ increases, the mass dimension of the Wilson coefficients decreases. In momentum space, this behaviour appears in the Fourier transformed version of Eq.~\eqref{eq:ope} as increasing powers of $1/Q^2$ in the $C_n$ constants. Therefore, condensates with higher dimensions are expected to have a small contribution to the OPE allowing the truncation of the OPE after a few non-perturbative terms.

The condensates are related to spontaneously broken symmetries, where the QCD Lagrangian exhibits a symmetry that it is not present in the QCD vacuum. The nonzero dimension 3 quark condensate ${\langle\Omega|:\bar{q}^{\phi}_k(0)q^{\phi}_k(0):|\Omega\rangle \equiv \langle\bar{q}q\rangle}$, for example, is responsible for the spontaneous breaking of chiral symmetry \cite{peskin,narison}, leading to the Gell-Mann-Oakes-Renner relation \cite{Gell-Mann:1968hlm}
\begin{align}
	\label{eq:pcac}
	m_q\langle\bar{q}q\rangle=-\frac{1}{2}f_{\pi}^2m_{\pi}^2\,,
\end{align}
where $m_q =\frac{1}{2}\left(m_u+m_d\right)$, $f_{\pi}=0.093$ GeV is the pion decay constant, and $m_{\pi}=0.139$ GeV is the pion mass \cite{pdg:2022}.

Consider the two-point correlation function in Eq.~\eqref{eq:correlator_composite} up to leading-order with the current in Eq.~\eqref{eq:heavy_light_current}. Applying the Wick's theorem we have:
{\allowdisplaybreaks\begin{equation}
		\label{eq:wick_theorem_diquark}
		\begin{split}
			\langle\Omega|TJ(x)J^{\dagger}(y)|\Omega\rangle &= \langle\Omega|:\contraction[2ex]{}{Q}{(x)q(x)\bar{q}(y)}{\bar{Q}}\contraction{Q(x)}{q}{(x)}{\bar{q}} Q(x)q(x)\bar{q}(y)\bar{Q}(y) + \contraction[2ex]{}{Q}{(x)q(x)\bar{q}(y)}{\bar{Q}}Q(x)q(x)\bar{q}(y)\bar{Q}(y) \\
			&+ \contraction{Q(x)}{q}{(x)}{\bar{q}}Q(x)q(x)\bar{q}(y)\bar{Q}(y) + Q(x)q(x)\bar{q}(y)\bar{Q}(y):|\Omega\rangle \\
			&= i^2 S^{(0)}_Q(x-y)S^{(0)}_q(x-y)\langle\Omega|:\mathbb{1}:|\Omega\rangle \\
			&- i S^{(0)}_Q(x-y) \langle\Omega|:\bar{q}(y)q(x):|\Omega\rangle \\
			&-i S^{(0)}_q(x-y) \langle\Omega|:\bar{Q}(y)Q(x):|\Omega\rangle \\
			&+ \langle\Omega|:Q(x)q(x)\bar{q}(y)\bar{Q}(y):|\Omega\rangle\,,
		\end{split}
\end{equation}}%
where colour indices, the Levi-Civita symbol and the charge conjugation operator were omitted for the sake of simplicity, and the minus signs are due to the anticommutation of the quark fields. The first term on the RHS of Eq.~(\ref{eq:wick_theorem_diquark}) is fully contracted and can be calculated using purely perturbation theory as in Eq.~\eqref{eq:correlator_expansion}, therefore, the first term represents the perturbative contribution to the correlation function. The second term in the RHS of Eq.~\eqref{eq:wick_theorem_diquark} is contracted only for the heavy quark-antiquark pair, resulting in a heavy-quark propagator $i S^{(0)}_Q(x-y)$, and in a nonvanishing light-quark vacuum expectation value $\langle\Omega|:\bar{q}(y)q(x):|\Omega\rangle$. The third term on the RHS of Eq.~\eqref{eq:wick_theorem_diquark} has the heavy quark-antiquark pair uncontracted, resulting in the heavy-quark VEV $\langle\Omega|:\bar{Q}(y)Q(x):|\Omega\rangle$ and in a light-quark propagator $i S^{(0)}_q(x-y)$. The fourth term on the RHS of Eq.~\eqref{eq:wick_theorem_diquark} is a fully uncontracted term and it does not contribute to the correlation function because it corresponds to a disconnected Feynman diagram, as seen in Section \ref{sec:two-point_correlation_function}. Hence, the second and third terms generate the condensates that represent non-perturbative contributions to the two-point correlation function.

The non-local vacuum expectation values in Eq.~\eqref{eq:wick_theorem_diquark} are then evaluated using the OPE, Eq.~\eqref{eq:ope}, where the Wilson coefficients can be calculated perturbatively using the fixed-point gauge method, and the properties of the QCD vacuum is also followed by the condensates, such as gauge and Lorentz invariance, resulting in \cite{Elias:PhysRevD.38.1584,Pascual:1984zb,Bagan:1994pw}
\begin{align}
	\begin{split}
		\label{eq:quark_vev_expansion}
		\langle\Omega|:&\bar{q}(y)^A_{\alpha, i}q^B_{\beta, j}(x):|\Omega\rangle = \frac{1}{2}\delta_{AB}\delta_{\alpha\beta}\Biggl\{ \left[ \delta_{ij} - \frac{i}{4} m_A(x-y)_{\mu} \gamma^{\mu}_{ji} \right] \langle\bar{q}q\rangle \\
		&+i\biggl[-\frac{i}{16}(x-y)^2\delta_{ij}-\frac{1}{24} \sigma_{ji}^{\mu\nu} x_{\mu}y_{\nu} - \frac{m_A}{96} \gamma^{\mu}_{ji} (x-y)_{\mu} (x-y)^2\biggr] \langle\bar{q}\sigma Gq\rangle \\
		&-\frac{i}{288}\biggl[g^2(x-y)^2 (x-y)_{\mu} \gamma^{\mu}_{ji}\biggr]\langle\bar{q}^A\gamma^{\rho}\frac{\lambda^a}{2} q^A\sum_C \bar{q}^C\gamma_{\rho}\frac{\lambda^a}{2}q^C\rangle\Biggr\} + \cdots \,,
	\end{split}    
\end{align}
where flavour, colour and Dirac indices are explicitly written, 
\begin{align}
	\langle\Omega|: g \bar{q}^{\alpha}_i(0) \sigma^{\mu\nu}_{ij} \frac{\lambda^a_{\alpha\beta}}{2} G^a_{\mu\nu}(0) q^{\beta}_j(0) :|\Omega\rangle \equiv \langle\bar{q}\sigma Gq\rangle
\end{align}
is the dimension 5 mixed condensate, $\sigma^{\mu\nu}=\frac{i}{2}\left[\gamma^{\mu},\gamma^{\nu}\right]$, 
\begin{align}
	\langle\bar{q}^A\gamma^{\rho}\frac{\lambda^a}{2} q^A\sum_C \bar{q}^C\gamma_{\rho}\frac{\lambda^a}{2}q^C\rangle = -\frac{4}{9}\langle\bar{q}q\rangle^2
	\label{eq:dim_6_condensate}
\end{align}
is the dimension 6 quark condensate, and the vacuum saturation hypothesis was used in Eq.~\eqref{eq:dim_6_condensate} \cite{Shifman:1978bx}.

The next two expansions are also relevant for the calculations of the heavy-light and heavy-strange diquarks in Chapter \ref{chapter:Paper_2} \cite{Pascual:1984zb,Elias:PhysRevD.38.1584}:
\begin{align}
	\begin{split}
		\label{eq:mix_vev_expansion}
		\langle\Omega|:&\bar{q}(x)_{\alpha, i}B^a_{\mu}(z)q_{\beta, j}(y):|\Omega\rangle = \frac{1}{768} z^{\omega_1} \Biggl\{ \biggl[ \frac{\lambda^a_{\beta\alpha}}{ig}\sigma^{\omega_1\mu}_{ij} \\
		&+ \frac{m_A}{2} \frac{\lambda^a_{\beta\alpha}}{ig}(y-x)^{\omega_2} \biggl(-i\sigma^{\omega_1\mu}_{ij}\gamma_{\omega_2}+g_{\omega_2\omega_1}\gamma_{\mu}-g_{\omega_2\mu}\gamma_{\omega_1}\biggr)_{ji}\biggr] i\langle\bar{q}\sigma Gq\rangle \\
		&+g^2\frac{\lambda^a_{\beta\alpha}}{ig}\biggl[\frac{2}{3}iz^{\omega_2}\biggl(-g_{\omega_1\omega_2}\gamma_{\mu}+ g_{\omega_2\mu}\gamma_{\omega_1}\biggr) + \frac{x^{\omega_2}}{2}\gamma_{\omega_2}\sigma^{\omega_1\mu} \\
		&- \frac{y^{\omega_2}}{2}\sigma^{\omega_1\mu}\gamma_{\omega_2}\biggr]_{ji}\langle\bar{q}^A\gamma^{\rho}\frac{\lambda^a}{2} q^A\sum_C \bar{q}^C\gamma_{\rho}\frac{\lambda^a}{2}q^C\rangle\Biggr\} + \cdots
	\end{split}    
\end{align}
and
\begin{align}
	\begin{split}
		\label{eq:gluon_vev_expansion}
		\langle\Omega|:&B^a_{\mu}(y)B^b_{\nu}(x):|\Omega\rangle = \frac{1}{32}\frac{1}{d(d-1)}y^{\rho}x^{\tau}\biggl[g_{\rho\tau}g_{\mu\nu}-g_{\rho\nu}-g_{\mu\tau}\biggr]\delta_{ab}\langle G^2\rangle + \cdots \,,
	\end{split}    
\end{align}
where 
\begin{align}
	\langle\Omega|:G^a_{\mu\nu}(0)G^a_{\mu\nu}(0):|\Omega\rangle \equiv \langle G^2\rangle
\end{align}
is the dimension 4 gluon condensate. Both non-local VEVs are generated when the higher order terms in Eq.~\eqref{eq:correlator_expansion} are included in the calculation.

It can be shown that the heavy-quark condensate is proportional to the gluon condensate \cite{Reinders:1984sr,narison}; therefore, to avoid overcounting of terms, the heavy quark condensate contribution can be omitted. Furthermore, the numerical value of the condensates are determined phenomenologically and enter as input parameters in Chapter \ref{chapter:Paper_2}.
\subsection{Hadronic Spectral Function}
\label{Sec:qcdsr_hadronic_spectral_function}
The hadronic spectral function $\rho^{\text{had}}(t)$ in Eq.~\eqref{eq:qcdsr_disp_rel} can sometimes be known experimentally, allowing the determination of QCD parameters such as condensates numerical values \cite{Shifman:1978by}, or it can be specified by a resonance model, allowing the prediction of hadronic properties such as diquark masses. In Chapter \ref{chapter:Paper_2} a single narrow resonance plus QCD continuum model is used for the hadronic spectral function,
\begin{align}
	\rho^{\text{had}}(t) = f^2\delta(t-M^2)+\theta(t-s_0)\frac{1}{\pi} \text{Im}\Pi(t) \,,
	\label{eq:single_resonance_model}
\end{align}
where $f \sim \langle\Omega|J|h\rangle$ indicates how strongly the bound state $|h\rangle$ is connected to the vacuum through the current $J$, $M$ is the predicted resonance ground state mass, $\theta(t-s_0)$ is the Heaviside step function, $s_0$ is the continuum threshold, and $\text{Im}\Pi(t)$ is the imaginary part of the two-point correlation function calculated in the QCD side of Eq.~\eqref{eq:qcdsr_disp_rel}, i.e., the two-point correlation function calculated using the OPE.
\subsection{Borel Transform}
\label{Sec:qcdsr_borel_transform}
In Eq.~\eqref{eq:qcdsr_disp_rel} the subtraction constants $\Pi^{(n)}(0)$ are polynomials in $Q^2$ and, usually, divergent. Due to the $Q^2$ polynomial structure of $\Pi^{(n)}(0)$, these divergences are called local divergences. In order to build a finite sum-rule that is able to predict diquark masses, as we shall see in Chapter \ref{chapter:Paper_2}, those potentially divergent subtraction constants need to be eliminated. One way of eliminating the subtraction constants is to apply the following Borel transform to both sides of Eq.~\eqref{eq:qcdsr_disp_rel} \cite{Pascual:1984zb}:
\begin{align}
	\label{eq:borel_transform}
	\mathcal{\hat B}=\lim_{\substack{N, Q^2\to \infty \\ \tau \equiv N/Q^2}}\frac{(-Q^2)^N}{\Gamma(N)}\left(\frac{d}{dQ^2}\right)^N \,,
\end{align}
where $\tau$ is the \textit{Borel parameter}. The \textit{Borel mass} is defined as $M_B = 1/\sqrt{\tau}$. Both parameters will be important to determine the stability and the optimization of the sum-rule in Section \ref{Sec:qcdsr_optimization}. 

From Eq.~\eqref{eq:borel_transform} the identities below can be shown:
\begin{subequations}
	\begin{align}
		&\mathcal{\hat B}\left[\frac{1}{(t+Q^2)^k}\right] = \frac{\tau^k}{\Gamma(k)}e^{-t\tau} \,, \label{eq:borel_identities_a} \\ \ \nonumber \\
		&\mathcal{\hat B}\left[\frac{Q^{2k}}{t+Q^2}\right] = \tau (-1)^k t^{k} e^{-t\tau} \,, \label{eq:borel_identities_b} \\ \ \nonumber \\
		&\mathcal{\hat B}\left[Q^{2k}\right] = 0 \,, \label{eq:borel_identities_c}
	\end{align}
\end{subequations}
where $k>0$. Notice that the Borel transform of a polynomial in $Q^2$ is zero (Eq.~\eqref{eq:borel_identities_c}), resulting in all local divergences being removed from both sides of the dispersion relation. Any non-local divergences, i.e., any divergences that are not a polynomial in $Q^2$, must be removed using a renormalization method. In Chaps.~\ref{chapter:Paper_1} and \ref{chapter:Paper_2} the diagrammatic renormalization method is used to deal with non-local divergences.
\subsection{Laplace Sum-Rules}
\label{Sec:qcdsr_laplace_sum_rules}
Applying the Borel transform identities \eqref{eq:borel_identities_b} and \eqref{eq:borel_identities_c}, and multiplying by $(-Q^2)^k$ on both sides of the dispersion relation \eqref{eq:qcdsr_disp_rel}, results in
\begin{align}
	\label{eq:qcdsr_disp_rel_borel_transform}
	\frac{\mathcal{\hat B}}{\tau}\left[(-Q^2)^k\Pi(Q^2)\right] = \int_{t_0}^{\infty} \rho^{\text{had}}(t) t^k e^{-t\tau} dt \,,
\end{align}
the Laplace sum-rules. The Borel transform \eqref{eq:borel_transform} is related to an inverse Laplace transform \cite{Bertlmann:1984ih,Pascual:1984zb} and therefore, the sum-rules built from the application of the Borel transform in the dispersion relation are called Laplace sum-rules.

Due to the exponential function in the integrand of Eq.~\eqref{eq:qcdsr_disp_rel_borel_transform}, the Borel transform results in sum-rules that suppress high-energy contributions. Therefore, Laplace sum-rules can be used to enhance the ground state of the hadronic spectral function.

Inserting Eq.~\eqref{eq:single_resonance_model} in Eq.~\eqref{eq:qcdsr_disp_rel_borel_transform} gives
\begin{align}
	\label{eq:qcdsr_r}
	\begin{split}
		\mathcal{R}_k(\tau,s_0) \equiv \frac{\mathcal{\hat B}}{\tau}\left[(-Q^2)^k\Pi(Q^2)\right] - \frac{1}{\pi}\int_{s_0}^{\infty} \text{Im}\Pi(t) t^k e^{-t\tau} dt = f^2 M^{2k} e^{-M^2\tau} \,,
	\end{split}
\end{align}
where $\Pi(Q^2)$ and $\text{Im}\Pi(t)$ are determined using the OPE, and the resonance mass can be extracted from the sum-rule by calculating the ratio,
\begin{align}
	\label{eq:qcdsr_mass}
	M(\tau,s_0) = \sqrt{\frac{\mathcal{R}_{k+1}(\tau,s_0)}{\mathcal{R}_k(\tau,s_0)}} \,.
\end{align}
Therefore, a hadronic mass parameter can be predicted from QCD calculations through $\mathcal{R}_k(\tau,s_0)$. In particular, the predicted resonance mass must be extracted from the possible $\tau$ and $s_0$ dependence on the RHS of Eq.~\eqref{eq:qcdsr_mass} through the various analysis procedures outlined in Section \ref{Sec:qcdsr_optimization}.
\subsection{Sum-Rules Optimization}
\label{Sec:qcdsr_optimization}
In Section \ref{Sec:qcdsr_hadronic_spectral_function} and \ref{Sec:qcdsr_borel_transform} the parameters $s_0$, the continuum threshold, and $\tau$, the Borel parameter were introduced to the QCDSR. Consequently, the hadronic mass extracted from the sum-rule in Eq.~\eqref{eq:qcdsr_mass} is a function of $\tau$ and $s_0$. Therefore, the determination of the parameters $\tau$ and $s_0$ is essential in order to properly predict hadronic properties from the QCDSR. This is done by considering some constraints on $\mathcal{R}_k(\tau,s_0)$ and $M(\tau,s_0)$.

\subsubsection{Borel Window}
Since the goal is to extract the resonance mass $M$ from the QCDSR, it is desired that the continuum contribution does not dominate the sum-rule. Furthermore, the OPE cannot be dominated by non-perturbative effects to guarantee its convergence. These two requirements set an upper and a lower bound for the Borel mass $M_B = 1/\sqrt{\tau}$, $M_B^{\text{max}}$ and $M_B^{\text{min}}$, respectively. The range $M_B^{\text{min}}\leq M_B \leq M_B^{\text{max}}$ is called \textit{Borel window}, and the sum-rule is considered reliable inside the Borel window \cite{Shifman:1978bx,Shifman:1978by}.

Following the same method as in Ref.~\cite{Kleiv:2013dta}, the upper bound on $M_B$ is determined considering that the contributions from the continuum are not more than $50\%$ of total contributions \cite{Shifman:1978by},
\begin{align}
	\label{eq:qcdsr_fcont}
	\frac{\mathcal{R}_1(\tau,s_0)/\mathcal{R}_0(\tau,s_0)}{\mathcal{R}_1(\tau,\infty)/\mathcal{R}_0(\tau,\infty)} \geq 0.5 \,,
\end{align}
and the lower bound on $M_B$ is determined by requiring that the sum-rule satisfy the H{\"o}lder inequality \cite{Benmerrouche:1995qa},
\begin{align}
	\label{eq:holder}
	\frac{\mathcal{R}_2(\tau,s_0)/\mathcal{R}_1(\tau,s_0)}{\mathcal{R}_1(\tau,s_0)/\mathcal{R}_0(\tau,s_0)} \geq 1 \,.
\end{align}
Note that no constrains were made at this point on $s_0$. Hence, there is a Borel window for each possible value of $s_0$.

\subsubsection{\texorpdfstring{$\tau$}{t}-Stability}
If the mass $M(\tau,s_0)$ satisfies
\begin{align}
	\label{eq:tau_stability}
	\frac{d}{d\tau}M(\tau,s_0)=0
\end{align}
within the Borel window, the sum-rule is said to exhibit $\tau$-stability. The requirement of $\tau$-stability leads to a constraint on the continuum threshold. The minimum value for the continuum threshold, $s_0^{\text{min}}$, is the smallest value of $s_0$ where Eq.~\eqref{eq:qcdsr_mass} satisfies Eq.~\eqref{eq:tau_stability} inside the Borel window.

\subsubsection{\texorpdfstring{$\chi^2$}{x} Minimization and \texorpdfstring{$s_0^{\text{opt}}$}{s0}}
Limiting the continuum contribution, and requiring that the sum-rule meet H{\"o}lder inequality and $\tau$-stability has determined $s_0^{\text{min}}$ and the Borel window for each value of $s_0$. The optimum value of the continuum threshold, $s_0^{\text{opt}}$, is determined by minimizing the $\chi^2(s_0)$ function
\begin{align}
	\label{eq:chi_squared}
	\chi^2(s_0)=\sum_{j=1}^{30}\left(\frac{1}{M}\sqrt{\frac{\mathcal{R}_1(\tau_j,s_0)}{\mathcal{R}_0(\tau_j,s_0)}}-1\right)^2 \,, \ s_0\geq s_0^{\text{min}} \,,
\end{align}
where the Borel window for each $s_0$, starting with $s_0^{\text{min}}$, is divided into 30 equally spaced points $\tau_j$, and $M$ is determined by a fit inside the Borel window. The optimum $s_0$ is the one where $\chi^2(s_0)$ has its lower value. The mass prediction is then given by fitting $M(\tau,s_0^{\text{opt}})$ to a constant in the $s_0^{\text{opt}}$ Borel window.

\chapter{Diagrammatic Renormalization Methods}
\label{chapter:Paper_1}
\section{Motivation and Background}
As discussed in Section \ref{sec:qcdsr_main_idea}, QCD sum-rules are based on quark-hadron duality to connect two-point correlation functions of composite operators calculated using the OPE to hadronic properties through the dispersion relation \eqref{eq:qcdsr_disp_rel}. In this process, all local divergences are eliminated by applying, for example, the Borel transform \eqref{eq:borel_transform}. As noted in Section \ref{Sec:qcdsr_borel_transform}, however, non-local divergences are not removed from the sum-rules with the application of the Borel transform, and a renormalization method needs to be used in order to deal with these divergences.

The conventional renormalization of composite operators, as outlined in Section \ref{Sec:qcd_regularization}, mixes operators, increasing the complexity of the renormalization process as the mass dimension of the operator increases \cite{Narison:1983kn,Jin:2000ek}. As shown in Ref.~\cite{deOliveira:2022eeq} and shown below, the diagrammatic renormalization method \cite{bogoliubov,Hepp:1966eg,Zimmermann:1969jj,Collins:1974da} (see, e.g., Refs.\cite{Collins:1984xc,muta} for a review), can be applied to QCD correlation functions to avoid those mixings, simplifying the renormalization of composite operators.

\begin{figure}[ht]
	\centering
	\includegraphics{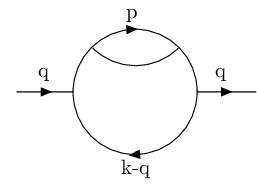}
	\caption{Two-loop contribution to the propagator in $\phi^3$ theory with external momentum $q$.}
	\label{fig:phi_cubed_NLO}
\end{figure}

In order to introduce the diagrammatic renormalization method, consider the $\phi^3$ theory
\begin{align}
	\label{eq:phi_cubed_lagrangian}
	\mathcal{L} = \frac{1}{2}(\partial_{\mu}\phi)^2 - \frac{1}{2}m_{\phi}^2\phi^2-\frac{\lambda}{3!}\phi^3 \,,
\end{align}
where $\phi(x)$ is a real scalar field, and $\lambda$ is a coupling. At two-loop order, one of the contributions to the scalar field self-energy is given by the Feynman diagram in Figure \ref{fig:phi_cubed_NLO}. A field theory is renormalizable when its coupling is dimensionless \cite{peskin}. Therefore, from a dimensional analysis (see Appendix \ref{Appendix_A}), the dimensions of the scalar field and the coupling are, respectively, $[\text{M}]^{\frac{1}{2}(d-2)}$ and $[\text{M}]^{\frac{1}{2}(6-d)}$, and the $\phi^3$ theory is renormalizable when $d=6$.

Applying dimensional regularization when $d=6+2\epsilon$ and setting the scalar field mass $m_{\phi}$ equals to zero, the contribution from Figure \ref{fig:phi_cubed_NLO} results in \cite{Collins:1984xc,muta}
{\allowdisplaybreaks\begin{align}
		\label{eq:phi_cubed_NLO}
		\begin{split}
			\Pi_{\text{bare}}(Q^2) &= i(-i\lambda)^4\mu^{2(6-d)}\frac{1}{2}\int\frac{d^dk d^dp}{(2\pi)^{2d}}\frac{i^5}{(k-q)^2(k)^2(p)^2(k-p)^2(k)^2} \\
			&=\frac{\lambda^4}{(4\pi)^6}\frac{Q^2}{144}\left\{-\frac{1}{\epsilon^2}-\frac{2}{\epsilon}\left[\log\left(\frac{Q^2}{\mu^2}\right)-\frac{43}{12}\right]\right\} + \mathcal{O}(\epsilon^0) \,,
		\end{split}
\end{align}}
where $Q^2=-q^2$, $\frac{1}{2}$ is a symmetry factor, and $\mu$ is the $\overline{\text{MS}}$-scheme renormalization scale. 

To renormalize the contribution from Figure \ref{fig:phi_cubed_NLO} using the diagrammatic renormalization method, first identify all its subdiagrams, i.e., all subsets of lines from the bare diagram that contains one or more loops. Second, extract the subdivergences from the subdiagrams, i.e., extract only the divergent part from the subdiagrams. Third, construct the counterterm diagram by replacing the subdiagram in the bare diagram by its subdivergence, where each subdiagram is associated with one counterterm. Finally, the renormalized diagram is calculated by subtracting all counterterm diagrams from the bare diagram. Any local divergences are removed by applying the Borel transform when building the QCD sum-rules, as discussed in Section \ref{sec:qcdsr_main_idea}.

\begin{figure}[ht]
	\centering
	\includegraphics{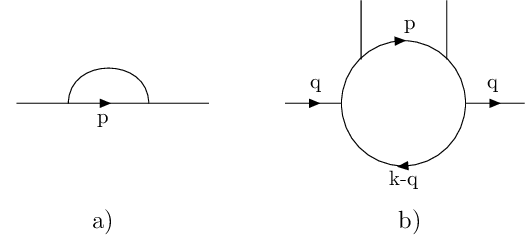}
	\caption{Subdiagrams extracted from the bare diagram in Figure \ref{fig:phi_cubed_NLO}.}
	\label{fig:phi_cubed_NLO_subdiagram}
\end{figure}

In Figure \ref{fig:phi_cubed_NLO} there are two subdiagrams, as shown in Figure \ref{fig:phi_cubed_NLO_subdiagram}. Contributions from these subdiagrams are,
\begin{align}
	\begin{split}
		\Sigma_a(k^2) &= (-i\lambda)^2\mu^{6-d}\int\frac{d^dp}{(2\pi)^d}\frac{i^2}{(k-p)^2(p)^2} \\
		&= -\frac{i\lambda^2}{(4\pi)^3}\frac{k^2}{6}\left[\frac{1}{\epsilon}-\log\left(\frac{-k^2}{\mu^2}\right)-\frac83\right] + \mathcal{O}(\epsilon^1) \,, \\ \ \\
		\Sigma_b(Q^2) &= (-i\lambda)^4\mu^{2(6-d)}\int\frac{d^dp}{(2\pi)^d}\frac{i^4}{(k-q)^2(k)^2(p)^2(k)^2}=0 \,.
	\end{split}
\end{align}
Therefore, subdiagram $a)$ contributes with a $-\frac{i\lambda^2}{(4\pi)^3}\frac{k^2}{6\epsilon}$ term, and subdiagram $b)$ contribution is zero.

\begin{figure}[ht]
	\centering
	\includegraphics{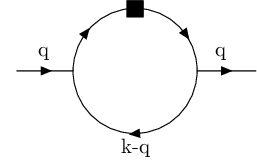}
	\caption{Counterterm diagram generated by subdiagram a) of Figure \ref{fig:phi_cubed_NLO_subdiagram}. The square $\blacksquare$ denotes the subdivergence insertion.}
	\label{fig:phi_cubed_counterterm}
\end{figure}

The counterterm diagram in constructed by replacing in the bare diagram in Figure \ref{fig:phi_cubed_NLO} the subdiagram $a)$ in Figure \ref{fig:phi_cubed_NLO_subdiagram} by its subdivergence, as shown in Figure \ref{fig:phi_cubed_counterterm}, where the square $\blacksquare$ represents the subdivergence contribution. The counterterm diagram contribution is
\begin{align}
	\begin{split}
		\Pi_{\text{ct}}(Q^2) &= i(-i\lambda)^2\mu^{6-d}\frac{1}{2}\left[-\frac{i\lambda^2}{(4\pi)^3}\frac{1}{6\epsilon}\right]\int\frac{d^dk}{(2\pi)^d}\frac{i^3 \ k^2}{(k-q)^2(k)^4} \\
		&=\frac{\lambda^4}{(4\pi)^6}\frac{Q^2}{144}\left\{-\frac{2}{\epsilon^2}-\frac{2}{\epsilon}\left[\log\left(\frac{Q^2}{\mu^2}\right)-\frac{8}{3}\right]\right\} + \mathcal{O}(\epsilon^0) \,,
	\end{split}
\end{align}
resulting in the renormalized diagram contribution below
\begin{align}
	\begin{split}
		\overline{\Pi}_{\text{rn}}(Q^2) &= \Pi_{\text{bare}}(Q^2) - \Pi_{\text{ct}}(Q^2) \\
		&= \frac{\lambda^4}{(4\pi)^6}\frac{Q^2}{144}\left\{\frac{1}{\epsilon^2}+\frac{2}{\epsilon}\frac{11}{12}\right\} + \mathcal{O}(\epsilon^0)\,.
	\end{split}
\end{align}
Note that $\overline{\Pi}_{\text{rn}}(Q^2)$ still have $1/\epsilon$ contributions. However, those divergences are local divergences (polynomials in $Q^2$), and can be removed by applying the Borel transform. Hence, the renormalized contribution $\Pi_{\text{rn}}(Q^2)$ is given by $\Pi_{\text{rn}}(Q^2) = \hat{\mathcal{B}}\left[\overline{\Pi}_{\text{rn}}(Q^2)\right]$.

Further applications of diagrammatic renormalization methods in QCD sum-rules for the scalar and vector mesonic correlation function for light and heavy quarks, heavy-light diquark correlation functions, and scalar quark meson glueball mixed correlation function are presented in Section \ref{sec:manuscript_1}.

\section{Applications of Diagrammatic Renormalization Methods in QCD Sum-Rules}	
\label{sec:manuscript_1}
This work may be found published in:
\begin{center}
	\noindent\fbox{%
		\parbox{25em}{
			Applications of Diagrammatic Renormalization Methods in QCD Sum-Rules
			
			T.~de Oliveira, D.~Harnett, A.~Palameta, and T.~G.~Steele
			
			Phys.~Rev.~D \textbf{106}, 114023 (2022)
			
			\href{https://journals.aps.org/prd/abstract/10.1103/PhysRevD.106.114023}{doi:10.1103/PhysRevD.106.114023}
		}%
	}
\end{center}

\section{Conclusion}
In this chapter the diagrammatic renormalization method is presented and applied in several examples in QCD sum-rules. In this method, differently from the conventional renormalization method where counterterms are added to the Lagrangian density of the theory being studied, individual Feynman diagrams are renormalized by identifying their subdiagrams, extracting the divergent part, constructing counterterm diagrams, and finally subtracting counterterm diagrams from bare diagrams. From mesonic to diquarks and mixed quark and glueball correlation functions, it was shown that the diagrammatic renormalization method is able to renormalize the correlation function and to agree with the conventional renormalization method result. In addition, with the diagrammatic renormalization method, the mixing of operators was avoided, resulting in improvements in the QCD sum-rule calculation for tetraquarks and pentaquarks, for example.

In Chapter \ref{chapter:Paper_2}, the diagrammatic renormalization method developed here is applied to the correlation function of heavy-light and doubly-strange diquarks.

\chapter{Heavy-Light and Doubly-Strange Diquarks}
\label{chapter:Paper_2}
\section{Motivation and Background}
As seen in Section \ref{Sec:BCQM_main_idea}, a tetraquark is a tightly-bound state of a coloured diquark and a coloured antidiquark. As a consequence of not being a colour singlet state, the diquark and the antidiquark are confined in the tetraquark and no free diquark or antidiquark has ever being measured. Diquark masses are an important parameter in tetraquark models (see, e.g., Ref.~\cite{Maiani:2004vq}), and can be extracted using QCD Laplace sum-rules where the QCD side is calculated using the OPE and the hadronic side is calculated using a single narrow resonance plus QCD continuum model.

In the manuscript below (see Section \ref{sec:manuscript_2}), $J^P\in\{0^{\pm},1^{\pm}\}$ heavy-light $[Qq]$, where $Q\in\{c,b\}$ and $q\in\{u,d,s\}$, and $J^P=1^+$ doubly-strange $[ss]$ diquarks are examined using QCD Laplace sum-rules. The two-point diquark correlation functions are renormalized using the diagrammatic renormalization method developed in Chapter \ref{chapter:Paper_1}, avoiding the mixing of composite operators. This work is an extension and an update of Ref.~\cite{Kleiv:2013dta}, where heavy-non-strange $[Qn]$ diquark masses are extracted from QCD Laplace sum-rules. The uncertainties in the parameters used in the analysis hide the heavy-strange $[Qs]$ and heavy-non-strange $[Qn]$ diquark mass splitting and, because of that, an analysis procedure is developed in the manuscript below, inspired by the double-ratio method in Ref.~\cite{Narison:1988ep}, in order to suppress the effects of the uncertainties from the input parameters. The strange quark condensate parameter $\kappa = \langle \bar{s} s \rangle/ \langle \bar{n}n\rangle$ is not suppressed in this methodology because it appears only in the strange channel. It is showed that $\kappa$ has an important role in determining the $[Qs]$ and $[Qn]$ mass splitting.

\subsection{Extension and Update}
\label{sec:extension and update}
Section \ref{sec:manuscript_2} is an extension and an update of Ref.~\cite{Kleiv:2013dta}. In this previous work, heavy-non-strange $[Qn]$ diquark masses are obtained from QCD Laplace sum-rules. In the manuscript below, heavy-non-strange $[Qn]$, heavy-strange $[Qs]$, and doubly-strange $[ss]$ diquark masses are calculated using QCD Laplace sum-rules. The diagrammatic renormalization method used in the manuscript agrees with the conventional renormalization method used in Ref.~\cite{Kleiv:2013dta}, resulting in another benchmark for the diagrammatic renormalization method. 

In the OPE calculation there are disagreements with the dimension-five and dimension-six condensate contributions. In Ref.~\cite{Kleiv:2013dta}, the result for the Feynman diagrams in Figures 8 b) and 9 b) does not agree with the manuscript result. The diagram in Figure 9 b) it is not considered at all in Ref.~\cite{Kleiv:2013dta}. These disagreements were resolved \cite{private_communication} and the updated results can be found in the manuscript. In addition, over the past decade, quark mass input parameters central values have changed and their uncertainties have improved, as shown in Table \ref{parameters_tab_comparison} \cite{pdg:2022,ParticleDataGroup:2012pjm}. All together, these improvements result in Table VI of the manuscript.

\begin{table}[hbt]
	\centering
	\renewcommand{\arraystretch}{1.5}
	\begin{tabular}{?c?c|c?c|c?}
		\thickhline
		\multirow{2}{*}{Parameter} & \multicolumn{2}{c?}{Charm} & \multicolumn{2}{c?}{Bottom}\\
		\hhline{|~|-|-|-|-|}
		& 2012 & 2022& 2012 & 2022 \\
		\thickhline
		$\overline{m}$ (GeV) &$1.28\pm0.03$& $1.27\pm0.02$ &$4.18\pm0.03$& $4.18\pm0.03$\\ \hline
		$r_{Qn}$ &$305\pm59$& $321.40\pm11.78$ &$1229\pm210$& $1474.18\pm44.81$\\ \thickhline
	\end{tabular}
	\caption{Comparison of 2012 and 2022 $\overline m ({\rm GeV})$ and $r_{Qn} = \frac{m ({\rm 2 GeV})}{m_n ({\rm 2 GeV})}$ input parameters, where $m$ is a heavy-quark mass.}
	\label{parameters_tab_comparison}
\end{table}

\section{Light-Quark $SU(3)$ Flavour Splitting of Heavy-Light Constituent Diquark Masses and Doubly-Strange Diquarks from QCD Sum-Rules}	
\label{sec:manuscript_2}

This work may be found published in:
\begin{center}
	\noindent\fbox{%
		\parbox{25em}{
			Light-Quark $SU(3)$ Flavour Splitting of Heavy-Light Constituent Diquark Masses and Doubly-Strange Diquarks from QCD Sum-Rules
			
			T.~de Oliveira, D.~Harnett, R.~Kleiv, A.~Palameta, and T.~G.~Steele
			
			Phys. Rev. D \textbf{108}, 054036 (2023)
			
			\href{https://journals.aps.org/prd/abstract/10.1103/PhysRevD.108.054036}{doi:10.1103/PhysRevD.108.054036}
		}%
	}
\end{center}

\section{Conclusion}
In Section \ref{sec:manuscript_2}, QCD Laplace sum-rules combined with diagrammatic renormalization methods were used to examine $J^P\in\{0^{\pm},1^{\pm}\}$ heavy-light $[Qq]$ and $J^P=1^+$ doubly-strange $[ss]$ diquarks. The heavy-light diquarks with negative parity and the doubly-strange diquark do not stabilize and, consequently, there is no mass prediction for these states. Despite the input parameters uncertainty improvements in the past decade discussed in Section \ref{sec:extension and update}, the $M_{[Qs]}-M_{[Qn]}$ mass splitting is hidden by the theoretical uncertainties and, as a result, an analysis methodology is developed in order to reduce these uncertainties. The strange quark condensate parameter $\kappa$ is not suppressed because it appears only in the strange channel, and it is showed that $\kappa$ has an important role in determining the mass splittings.

\begin{figure}[hbt]
	\centering
	\begin{tabular}{cc}
		\includegraphics[width=0.45\textwidth]{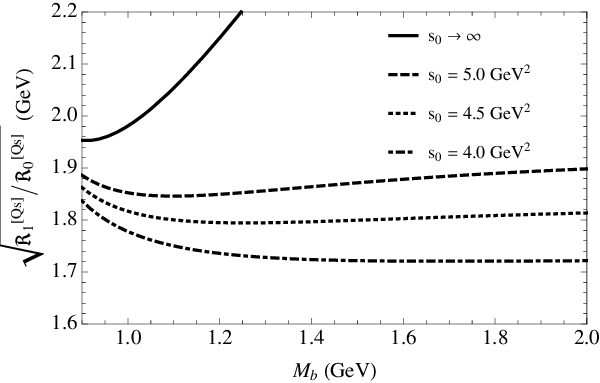}&  \includegraphics[width=0.45\textwidth]{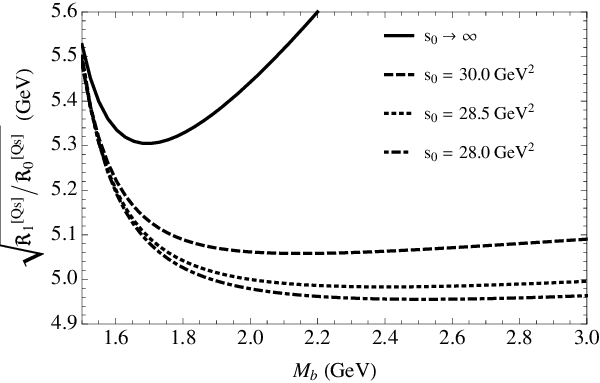}\\
		\includegraphics[width=0.45\textwidth]{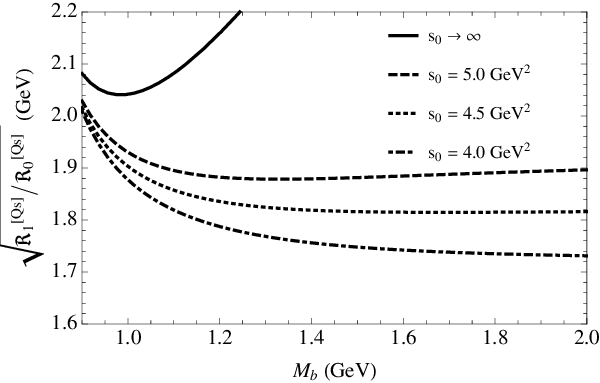}& \includegraphics[width=0.45\textwidth]{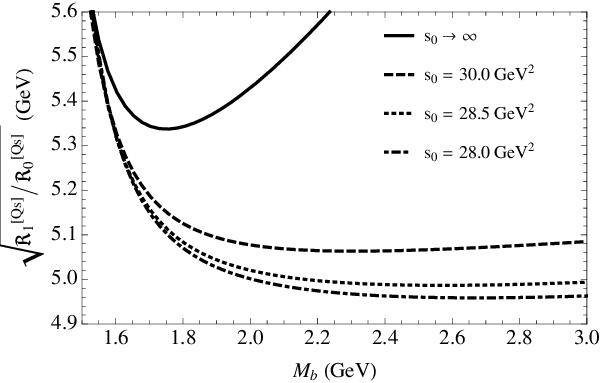}
	\end{tabular}
	\caption{The quantity $\sqrt{\frac{\mathcal{R}_1^{[Qs]}(M_b, s_0)}{\mathcal{R}_0^{[Qs]}(M_b, s_0)}}$ for $0^+$ charm-strange (top left), $0^+$ bottom-strange (top right), $1^+$ charm-strange (bottom left), and $1^+$ bottom-strange (bottom right) diquarks for selected values of $s_0$ and with $\kappa=0.74$.}
	\label{fig:r1r0_Qs}
\end{figure}

\begin{figure}[hbt]
	\centering
	\begin{tabular}{cc}
		\includegraphics[width=0.45\textwidth]{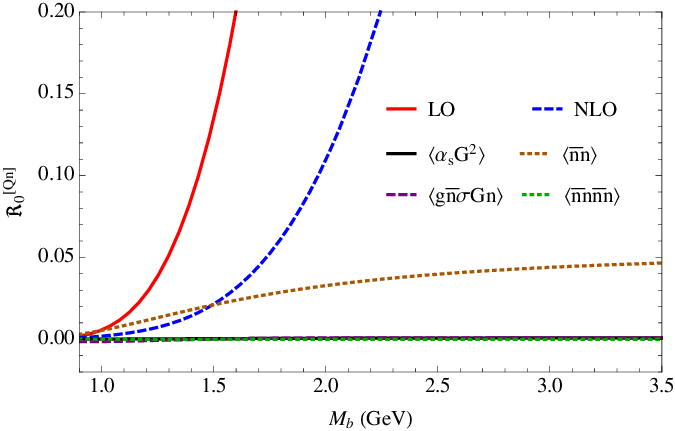}&  \includegraphics[width=0.45\textwidth]{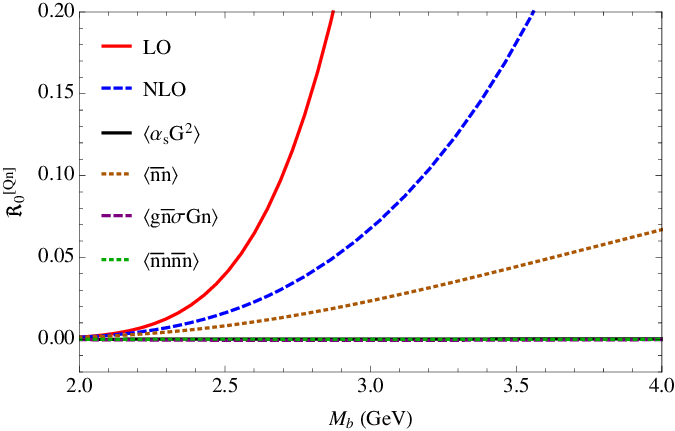}\\
		\includegraphics[width=0.45\textwidth]{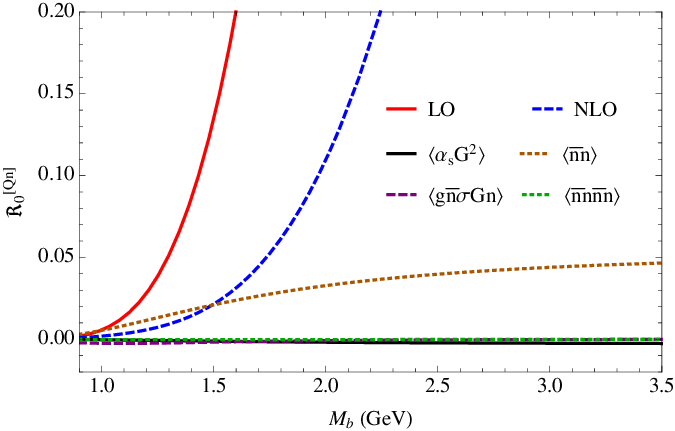}& \includegraphics[width=0.45\textwidth]{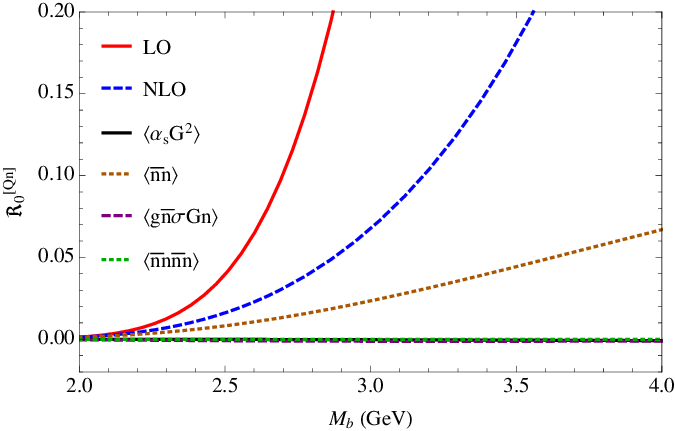}
	\end{tabular}
	\caption{$\mathcal{R}_0^{[Qn]}(M_b, s_0)$ contributions for $0^+$ charm (top left), $0^+$ bottom (top right), $1^+$ charm (bottom left), and $1^+$ bottom (bottom right) diquarks for $s_0\rightarrow\infty$.}
	\label{fig:r0_Qn}
\end{figure}

\begin{figure}[hbt]
	\centering
	\begin{tabular}{cc}
		\includegraphics[width=0.45\textwidth]{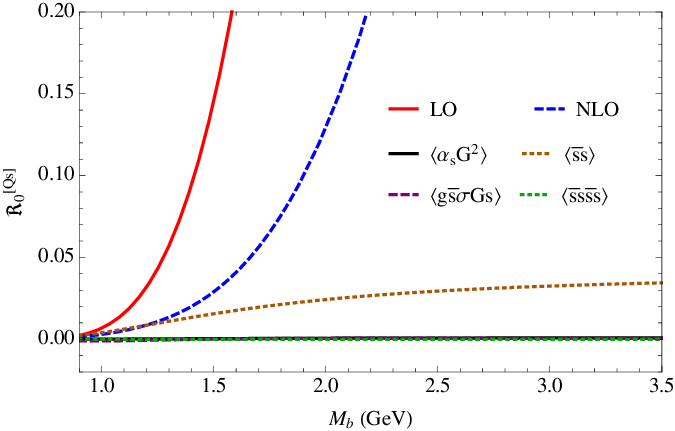}&  \includegraphics[width=0.45\textwidth]{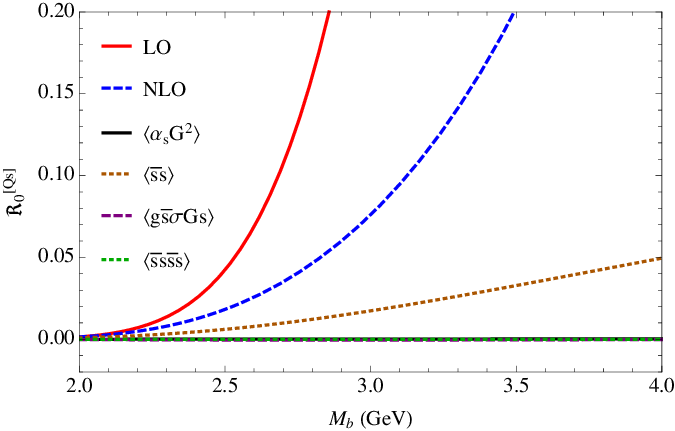}\\
		\includegraphics[width=0.45\textwidth]{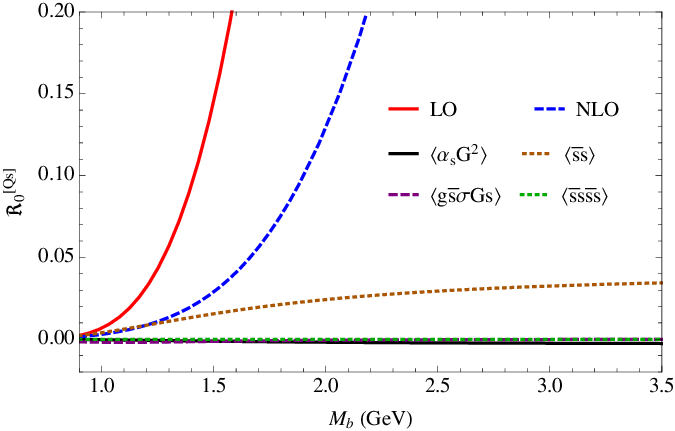}& \includegraphics[width=0.45\textwidth]{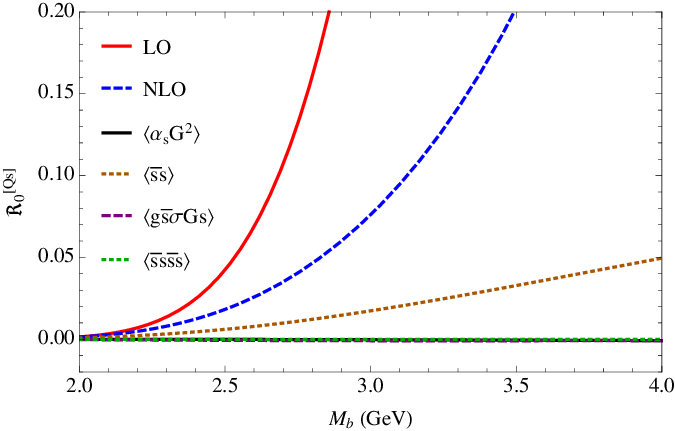}
	\end{tabular}
	\caption{$\mathcal{R}_0^{[Qs]}(M_b, s_0)$ contributions for $0^+$ charm (top left), $0^+$ bottom (top right), $1^+$ charm (bottom left), and $1^+$ bottom (bottom right) diquarks for $s_0\rightarrow\infty$ and with $\kappa=0.74$.}
	\label{fig:r0_Qs}
\end{figure}

In addition to the manuscript presented in Section \ref{sec:manuscript_2}, Figure \ref{fig:r1r0_Qs} shows the quantity $\sqrt{\frac{\mathcal{R}_1^{[Qs]}(M_b, s_0)}{\mathcal{R}_0^{[Qs]}(M_b, s_0)}}$ only for $[Qs]$ diquarks for $\kappa=0.74$ and $s_0$ near the $[Qn]$ $s_0^{\rm opt}$. Figures \ref{fig:r0_Qn} and \ref{fig:r0_Qs} show the contribution to $\mathcal{R}_0^{[Qq]}(M_b, s_0)$ from the leading-order (LO), next-to-leading-order (NLO), dimension-three, dimension-four, dimension-five, and dimension-six condensates. After the perturbative contribution (LO and NLO), the dimension-three quark condensate is the most important contribution to $\mathcal{R}_0^{[Qq]}(M_b, s_0)$ in all cases (similar figures are found for $\mathcal{R}_1^{[Qq]}(M_b, s_0)$). The dimension-three quark condensate is proportional to $r_{Qn}$, therefore, $r_{Qn}$ is the next most important parameter after $\kappa$. The analysis of the uncertainty in $r_{Qn}$ results in Table \ref{tab:rQ}. As stated in the manuscript, the uncertainty in $r_{Qn}$ results in $\sim 5\,{\rm MeV}$ uncertainty in the $M_{[Qs]}-M_{[Qn]}$ mass splittings, therefore, the theoretical uncertainty in $\kappa$ is the dominant effect.

\begin{table}[hbt]
	\centering
	\renewcommand{\arraystretch}{1.5}
	\begin{tabular}{?c|c|c|c|c|c|c|c?}
		\thickhline
		\multirow{ 2}{*}{$[Qq]$} & \multirow{ 2}{*}{$J^P$} & \multirow{ 2}{*}{$s_0^{\mathrm{opt}}$ ($\mathrm{GeV}^2$)} & \multirow{ 2}{*}{$M_{Qq}$ (GeV)} & \multicolumn{2}{c|}{$\Delta$ (MeV)} & \multicolumn{2}{c?}{$M_{Qs}^{\Delta}$ (GeV)}\\
		\hhline{|~|~|~|~|-|-|-|-|} &&&&$r_Q^{min}$&$r_Q^{max}$&$r_Q^{min}$&$r_Q^{max}$\\ \thickhline
		\multirow{ 2}{*}{$[cq]$} & $0^+$& 4.50 & 1.78 & 51.1 & 57.5 & 1.83 & 1.84\\
		& $1^+$ & 5.00 & 1.87 & 52.2 & 60.2 & 1.92 & 1.93\\ \thickhline
		\multirow{ 2}{*}{$[bq]$} & $0^+$& 28.5 & 4.97 & 74.2 & 79.9 & 5.04 & 5.05\\
		& $1^+$ & 28.5 & 4.97 & 85.0 & 92.0 & 5.05 & 5.06\\ \thickhline       
	\end{tabular}
	\caption{Uncertainty analysis for $r_{Qn}$ with $\kappa=0.74$.}
	\label{tab:rQ}
\end{table}

\chapter{Conclusion}
\label{chapter:Conclusion}
In the past two decades many exotic hadron candidates have been detected (see e.g., Refs.~\cite{Swanson:2006st,Godfrey:2008nc,Olsen:RevModPhys.90.015003,Brambilla:2019esw} for reviews). The hadronic content of exotic hadrons varies from glueballs to multiquark bound states. In particular, the internal structure of four-quark states can be divided into two types: molecules and tetraquarks. Molecules are loosely bound with large-distances interactions described by one-pion exchange. Tetraquarks are tightly bound through the colour force and are composed of a diquark and an antidiquark. The main focus of this thesis was to explore diquarks in order to build knowledge for current and future experiments and models.

Diquark (and hadronic) properties can be extracted from QCD calculations by applying the QCD sum-rule method. In this method, the diquark two-point correlation function is connected to a spectral function through a dispersion relation. In this process, a renormalization method is needed so that non-local divergences are removed from the two-point correlation function. The renormalization of correlation functions of composite operators in the conventional renormalization method increases in complexity as the mass dimension of the operators increases because of the mixing of operators. Therefore, in the manuscript presented in Chapter \ref{chapter:Paper_1}, to avoid the mixing of operators, the diagrammatic renormalization method was applied to many examples of correlation functions in the context of QCD Laplace sum-rules. It was shown that the diagrammatic renormalization method agrees with the conventional renormalization method, and that the mixing of operators is avoided. In future work, the diagrammatic renormalization method represents an increase in the efficiency of the renormalization of two-point correlation functions of composite operators for QCD sum-rules calculations.

In Chapter \ref{chapter:Paper_2}, QCD Laplace sum-rules were used to examine $J^P\in\{0^{\pm},1^{\pm}\}$ heavy-light $[Qq]$ and $J^P=1^+$ doubly-strange $[ss]$ diquarks. The two-point correlation functions were renormalized using the diagrammatic renormalization method developed in Chapter \ref{chapter:Paper_1}. Only the sum-rule for the $J^P\in\{0^+,1^+\}$ heavy-light $[Qq]$ diquarks exhibits $\tau$-stability, allowing the extraction of a mass prediction. The mass splitting between the heavy-strange $[Qs]$ and the heavy-non-strange $[Qn]$ diquark, however, is obscured by the theoretical uncertainties of the input parameters. An analysis methodology was developed in the manuscript, inspired by the double-ratio method \cite{Narison:1988ep}, to suppress the theoretical uncertainties from all input parameters except the strange quark condensate parameter $\kappa = \frac{\langle\bar{s}s\rangle}{\langle\bar{n}n\rangle}$. As a result, it was shown that for $0.56 < \kappa < 0.74$, $55\,{\rm MeV} \lesssim M_{[cs]}-M_{[cn]}\lesssim 100\,{\rm MeV}$ and $75\,{\rm MeV} \lesssim M_{[bs]}-M_{[bn]}\lesssim 150\,{\rm MeV}$, in agreement with the constituent diquark mass parameters used in many models for tetraquarks, as noted in Section \ref{sec:manuscript_2}, providing supporting QCD evidence for the diquark constituent mass parameters used in these models. In addition, the results found and techniques developed in Chapter \ref{chapter:Paper_2} can improve future work on diquarks and can guide interpretations of the internal structure of tetraquarks. Having established diagrammatic renormalization as a valuable methodology for QCD sum-rule studies of diquarks,  this thesis opens the possibility of future studies of more challenging systems such as $[bc]$ diquarks.

\addcontentsline{toc}{chapter}{References}
\renewcommand\bibname{References}
\bibliography{99Biblio} 

\begin{thebibliography}{71}%
\makeatletter
\providecommand \@ifxundefined [1]{%
 \@ifx{#1\undefined}
}%
\providecommand \@ifnum [1]{%
 \ifnum #1\expandafter \@firstoftwo
 \else \expandafter \@secondoftwo
 \fi
}%
\providecommand \@ifx [1]{%
 \ifx #1\expandafter \@firstoftwo
 \else \expandafter \@secondoftwo
 \fi
}%
\providecommand \natexlab [1]{#1}%
\providecommand \enquote  [1]{``#1''}%
\providecommand \bibnamefont  [1]{#1}%
\providecommand \bibfnamefont [1]{#1}%
\providecommand \citenamefont [1]{#1}%
\providecommand \href@noop [0]{\@secondoftwo}%
\providecommand \href [0]{\begingroup \@sanitize@url \@href}%
\providecommand \@href[1]{\@@startlink{#1}\@@href}%
\providecommand \@@href[1]{\endgroup#1\@@endlink}%
\providecommand \@sanitize@url [0]{\catcode `\\12\catcode `\$12\catcode
  `\&12\catcode `\#12\catcode `\^12\catcode `\_12\catcode `\%12\relax}%
\providecommand \@@startlink[1]{}%
\providecommand \@@endlink[0]{}%
\providecommand \url  [0]{\begingroup\@sanitize@url \@url }%
\providecommand \@url [1]{\endgroup\@href {#1}{\urlprefix }}%
\providecommand \urlprefix  [0]{URL }%
\providecommand \Eprint [0]{\href }%
\providecommand \doibase [0]{https://doi.org/}%
\providecommand \selectlanguage [0]{\@gobble}%
\providecommand \bibinfo  [0]{\@secondoftwo}%
\providecommand \bibfield  [0]{\@secondoftwo}%
\providecommand \translation [1]{[#1]}%
\providecommand \BibitemOpen [0]{}%
\providecommand \bibitemStop [0]{}%
\providecommand \bibitemNoStop [0]{.\EOS\space}%
\providecommand \EOS [0]{\spacefactor3000\relax}%
\providecommand \BibitemShut  [1]{\csname bibitem#1\endcsname}%
\let\auto@bib@innerbib\@empty
%</preamble>
\bibitem [{\citenamefont {Choi}\ \emph {et~al.}(2003)\citenamefont {Choi} \emph
  {et~al.}}]{X3872:PhysRevLett.91.262001}%
  \BibitemOpen
  \bibfield  {author} {\bibinfo {author} {\bibfnamefont {S.~K.}\ \bibnamefont
  {Choi}} \emph {et~al.} (\bibinfo {collaboration} {Belle}),\ }\href
  {https://doi.org/10.1103/PhysRevLett.91.262001} {\bibfield  {journal}
  {\bibinfo  {journal} {Phys. Rev. Lett.}\ }\textbf {\bibinfo {volume} {91}},\
  \bibinfo {pages} {262001} (\bibinfo {year} {2003})},\ \Eprint
  {https://arxiv.org/abs/hep-ex/0309032} {arXiv:hep-ex/0309032} \BibitemShut
  {NoStop}%
\bibitem [{\citenamefont {Swanson}(2006)}]{Swanson:2006st}%
  \BibitemOpen
  \bibfield  {author} {\bibinfo {author} {\bibfnamefont {E.~S.}\ \bibnamefont
  {Swanson}},\ }\href {https://doi.org/10.1016/j.physrep.2006.04.003}
  {\bibfield  {journal} {\bibinfo  {journal} {Phys. Rept.}\ }\textbf {\bibinfo
  {volume} {429}},\ \bibinfo {pages} {243} (\bibinfo {year} {2006})},\ \Eprint
  {https://arxiv.org/abs/hep-ph/0601110} {arXiv:hep-ph/0601110} \BibitemShut
  {NoStop}%
\bibitem [{\citenamefont {Godfrey}\ and\ \citenamefont
  {Olsen}(2008)}]{Godfrey:2008nc}%
  \BibitemOpen
  \bibfield  {author} {\bibinfo {author} {\bibfnamefont {S.}~\bibnamefont
  {Godfrey}}\ and\ \bibinfo {author} {\bibfnamefont {S.~L.}\ \bibnamefont
  {Olsen}},\ }\href {https://doi.org/10.1146/annurev.nucl.58.110707.171145}
  {\bibfield  {journal} {\bibinfo  {journal} {Ann. Rev. Nucl. Part. Sci.}\
  }\textbf {\bibinfo {volume} {58}},\ \bibinfo {pages} {51} (\bibinfo {year}
  {2008})},\ \Eprint {https://arxiv.org/abs/0801.3867} {arXiv:0801.3867
  [hep-ph]} \BibitemShut {NoStop}%
\bibitem [{\citenamefont {Olsen}\ \emph {et~al.}(2018)\citenamefont {Olsen},
  \citenamefont {Skwarnicki},\ and\ \citenamefont
  {Zieminska}}]{Olsen:RevModPhys.90.015003}%
  \BibitemOpen
  \bibfield  {author} {\bibinfo {author} {\bibfnamefont {S.~L.}\ \bibnamefont
  {Olsen}}, \bibinfo {author} {\bibfnamefont {T.}~\bibnamefont {Skwarnicki}},\
  and\ \bibinfo {author} {\bibfnamefont {D.}~\bibnamefont {Zieminska}},\ }\href
  {https://doi.org/10.1103/RevModPhys.90.015003} {\bibfield  {journal}
  {\bibinfo  {journal} {Rev. Mod. Phys.}\ }\textbf {\bibinfo {volume} {90}},\
  \bibinfo {pages} {015003} (\bibinfo {year} {2018})},\ \Eprint
  {https://arxiv.org/abs/1708.04012} {arXiv:1708.04012 [hep-ph]} \BibitemShut
  {NoStop}%
\bibitem [{\citenamefont {Brambilla}\ \emph {et~al.}(2020)\citenamefont
  {Brambilla}, \citenamefont {Eidelman}, \citenamefont {Hanhart}, \citenamefont
  {Nefediev}, \citenamefont {Shen}, \citenamefont {Thomas}, \citenamefont
  {Vairo},\ and\ \citenamefont {Yuan}}]{Brambilla:2019esw}%
  \BibitemOpen
  \bibfield  {author} {\bibinfo {author} {\bibfnamefont {N.}~\bibnamefont
  {Brambilla}}, \bibinfo {author} {\bibfnamefont {S.}~\bibnamefont {Eidelman}},
  \bibinfo {author} {\bibfnamefont {C.}~\bibnamefont {Hanhart}}, \bibinfo
  {author} {\bibfnamefont {A.}~\bibnamefont {Nefediev}}, \bibinfo {author}
  {\bibfnamefont {C.-P.}\ \bibnamefont {Shen}}, \bibinfo {author}
  {\bibfnamefont {C.~E.}\ \bibnamefont {Thomas}}, \bibinfo {author}
  {\bibfnamefont {A.}~\bibnamefont {Vairo}},\ and\ \bibinfo {author}
  {\bibfnamefont {C.-Z.}\ \bibnamefont {Yuan}},\ }\href
  {https://doi.org/10.1016/j.physrep.2020.05.001} {\bibfield  {journal}
  {\bibinfo  {journal} {Phys. Rept.}\ }\textbf {\bibinfo {volume} {873}},\
  \bibinfo {pages} {1} (\bibinfo {year} {2020})},\ \Eprint
  {https://arxiv.org/abs/1907.07583} {arXiv:1907.07583 [hep-ex]} \BibitemShut
  {NoStop}%
\bibitem [{\citenamefont {Aaij}\ \emph {et~al.}(2022)\citenamefont {Aaij} \emph
  {et~al.}}]{LHCb:2022xob}%
  \BibitemOpen
  \bibfield  {author} {\bibinfo {author} {\bibfnamefont {R.}~\bibnamefont
  {Aaij}} \emph {et~al.} (\bibinfo {collaboration} {LHCb}),\ }\Eprint
  {https://arxiv.org/abs/2212.02716} {arXiv:2212.02716 [hep-ex]}  (\bibinfo
  {year} {2022}),\ \bibinfo {note} {{submitted to Phys. Rev.
  Lett.}}\BibitemShut {Stop}%
\bibitem [{\citenamefont {Maiani}\ \emph {et~al.}(2005)\citenamefont {Maiani},
  \citenamefont {Piccinini}, \citenamefont {Polosa},\ and\ \citenamefont
  {Riquer}}]{Maiani:2004vq}%
  \BibitemOpen
  \bibfield  {author} {\bibinfo {author} {\bibfnamefont {L.}~\bibnamefont
  {Maiani}}, \bibinfo {author} {\bibfnamefont {F.}~\bibnamefont {Piccinini}},
  \bibinfo {author} {\bibfnamefont {A.~D.}\ \bibnamefont {Polosa}},\ and\
  \bibinfo {author} {\bibfnamefont {V.}~\bibnamefont {Riquer}},\ }\href
  {https://doi.org/10.1103/PhysRevD.71.014028} {\bibfield  {journal} {\bibinfo
  {journal} {Phys. Rev. D}\ }\textbf {\bibinfo {volume} {71}},\ \bibinfo
  {pages} {014028} (\bibinfo {year} {2005})},\ \Eprint
  {https://arxiv.org/abs/hep-ph/0412098} {arXiv:hep-ph/0412098} \BibitemShut
  {NoStop}%
\bibitem [{\citenamefont {Aad}\ \emph {et~al.}(2012)\citenamefont {Aad} \emph
  {et~al.}}]{Higgs1:20121}%
  \BibitemOpen
  \bibfield  {author} {\bibinfo {author} {\bibfnamefont {G.}~\bibnamefont
  {Aad}} \emph {et~al.} (\bibinfo {collaboration} {ATLAS}),\ }\href
  {https://doi.org/10.1016/j.physletb.2012.08.020} {\bibfield  {journal}
  {\bibinfo  {journal} {Phys. Lett. B}\ }\textbf {\bibinfo {volume} {716}},\
  \bibinfo {pages} {1} (\bibinfo {year} {2012})},\ \Eprint
  {https://arxiv.org/abs/1207.7214} {arXiv:1207.7214 [hep-ex]} \BibitemShut
  {NoStop}%
\bibitem [{\citenamefont {Chatrchyan}\ \emph {et~al.}(2012)\citenamefont
  {Chatrchyan} \emph {et~al.}}]{Higgs2:201230}%
  \BibitemOpen
  \bibfield  {author} {\bibinfo {author} {\bibfnamefont {S.}~\bibnamefont
  {Chatrchyan}} \emph {et~al.} (\bibinfo {collaboration} {CMS}),\ }\href
  {https://doi.org/10.1016/j.physletb.2012.08.021} {\bibfield  {journal}
  {\bibinfo  {journal} {Phys. Lett. B}\ }\textbf {\bibinfo {volume} {716}},\
  \bibinfo {pages} {30} (\bibinfo {year} {2012})},\ \Eprint
  {https://arxiv.org/abs/1207.7235} {arXiv:1207.7235 [hep-ex]} \BibitemShut
  {NoStop}%
\bibitem [{\citenamefont {Workman}\ \emph {et~al.}(2022)\citenamefont {Workman}
  \emph {et~al.}}]{pdg:2022}%
  \BibitemOpen
  \bibfield  {author} {\bibinfo {author} {\bibfnamefont {R.~L.}\ \bibnamefont
  {Workman}} \emph {et~al.} (\bibinfo {collaboration} {Particle Data Group}),\
  }\href {https://doi.org/10.1093/ptep/ptac097} {\bibfield  {journal} {\bibinfo
   {journal} {PTEP}\ }\textbf {\bibinfo {volume} {2022}},\ \bibinfo {pages}
  {083C01} (\bibinfo {year} {2022})}\BibitemShut {NoStop}%
\bibitem [{\citenamefont {Gell-Mann}(1964)}]{Gell-Mann:1964ewy}%
  \BibitemOpen
  \bibfield  {author} {\bibinfo {author} {\bibfnamefont {M.}~\bibnamefont
  {Gell-Mann}},\ }\href {https://doi.org/10.1016/S0031-9163(64)92001-3}
  {\bibfield  {journal} {\bibinfo  {journal} {Phys. Lett.}\ }\textbf {\bibinfo
  {volume} {8}},\ \bibinfo {pages} {214} (\bibinfo {year} {1964})}\BibitemShut
  {NoStop}%
\bibitem [{\citenamefont {Zweig}(1964)}]{Zweig:1964jf}%
  \BibitemOpen
  \bibfield  {author} {\bibinfo {author} {\bibfnamefont {G.}~\bibnamefont
  {Zweig}},\ }\bibinfo {title} {{An SU(3) model for strong interaction symmetry
  and its breaking. Version 2}},\ in\ \href@noop {} {\emph {\bibinfo
  {booktitle} {{Developments in the Quark Theory of Hadrons. Vol. 1.}}}},\
  \bibinfo {editor} {edited by\ \bibinfo {editor} {\bibfnamefont {D.~B.}\
  \bibnamefont {Lichtenberg}}\ and\ \bibinfo {editor} {\bibfnamefont {S.~P.}\
  \bibnamefont {Rosen}}}\ (\bibinfo  {publisher} {Hadronic Press},\ \bibinfo
  {year} {1964})\ pp.\ \bibinfo {pages} {22--101}\BibitemShut {NoStop}%
\bibitem [{\citenamefont {Gell-Mann}(1961)}]{Gell-Mann:1961eightfold}%
  \BibitemOpen
  \bibfield  {author} {\bibinfo {author} {\bibfnamefont {M.}~\bibnamefont
  {Gell-Mann}},\ }\bibfield  {journal} {\bibinfo  {journal} {CTSL-20}\ }\textbf
  {\bibinfo {volume} {TID-12608}},\ \href {https://doi.org/10.2172/4008239}
  {10.2172/4008239} (\bibinfo {year} {1961})\BibitemShut {NoStop}%
\bibitem [{\citenamefont {Ne'eman}(1961)}]{Neeman:1961eightfold}%
  \BibitemOpen
  \bibfield  {author} {\bibinfo {author} {\bibfnamefont {Y.}~\bibnamefont
  {Ne'eman}},\ }\href {https://doi.org/10.1016/0029-5582(61)90134-1} {\bibfield
   {journal} {\bibinfo  {journal} {Nucl. Phys.}\ }\textbf {\bibinfo {volume}
  {26}},\ \bibinfo {pages} {222} (\bibinfo {year} {1961})}\BibitemShut
  {NoStop}%
\bibitem [{\citenamefont {Greenberg}(1964)}]{Greenberg:1964pe}%
  \BibitemOpen
  \bibfield  {author} {\bibinfo {author} {\bibfnamefont {O.~W.}\ \bibnamefont
  {Greenberg}},\ }\href {https://doi.org/10.1103/PhysRevLett.13.598} {\bibfield
   {journal} {\bibinfo  {journal} {Phys. Rev. Lett.}\ }\textbf {\bibinfo
  {volume} {13}},\ \bibinfo {pages} {598} (\bibinfo {year} {1964})}\BibitemShut
  {NoStop}%
\bibitem [{\citenamefont {Narison}(2004)}]{narison}%
  \BibitemOpen
  \bibfield  {author} {\bibinfo {author} {\bibfnamefont {S.}~\bibnamefont
  {Narison}},\ }\href {https://doi.org/10.1017/CBO9780511535000} {\emph
  {\bibinfo {title} {QCD as a Theory of Hadrons: From Partons to
  Confinement}}},\ Cambridge Monographs on Particle Physics, Nuclear Physics
  and Cosmology\ (\bibinfo  {publisher} {Cambridge University Press},\ \bibinfo
  {year} {2004})\BibitemShut {NoStop}%
\bibitem [{\citenamefont {Feger}\ \emph {et~al.}(2020)\citenamefont {Feger},
  \citenamefont {Kephart},\ and\ \citenamefont {Saskowski}}]{Feger:2019tvk}%
  \BibitemOpen
  \bibfield  {author} {\bibinfo {author} {\bibfnamefont {R.}~\bibnamefont
  {Feger}}, \bibinfo {author} {\bibfnamefont {T.~W.}\ \bibnamefont {Kephart}},\
  and\ \bibinfo {author} {\bibfnamefont {R.~J.}\ \bibnamefont {Saskowski}},\
  }\href {https://doi.org/10.1016/j.cpc.2020.107490} {\bibfield  {journal}
  {\bibinfo  {journal} {Comput. Phys. Commun.}\ }\textbf {\bibinfo {volume}
  {257}},\ \bibinfo {pages} {107490} (\bibinfo {year} {2020})},\ \Eprint
  {https://arxiv.org/abs/1912.10969} {arXiv:1912.10969 [hep-th]} \BibitemShut
  {NoStop}%
\bibitem [{\citenamefont {Cheng}\ and\ \citenamefont
  {Li}(1984)}]{Cheng:1984vwu}%
  \BibitemOpen
  \bibfield  {author} {\bibinfo {author} {\bibfnamefont {T.-P.}\ \bibnamefont
  {Cheng}}\ and\ \bibinfo {author} {\bibfnamefont {L.-F.}\ \bibnamefont {Li}},\
  }\href@noop {} {\emph {\bibinfo {title} {{Gauge Theory of Elementary Particle
  Physics}}}}\ (\bibinfo  {publisher} {Oxford University Press},\ \bibinfo
  {address} {Oxford, UK},\ \bibinfo {year} {1984})\BibitemShut {NoStop}%
\bibitem [{\citenamefont
  {Olsen}(2015{\natexlab{a}})}]{Olsen:newhadronspectroscopy2015}%
  \BibitemOpen
  \bibfield  {author} {\bibinfo {author} {\bibfnamefont {S.~L.}\ \bibnamefont
  {Olsen}},\ }\href {https://doi.org/10.1007/S11467-014-0449-6} {\bibfield
  {journal} {\bibinfo  {journal} {Front. Phys. (Beijing)}\ }\textbf {\bibinfo
  {volume} {10}},\ \bibinfo {pages} {121} (\bibinfo {year}
  {2015}{\natexlab{a}})},\ \Eprint {https://arxiv.org/abs/1411.7738}
  {arXiv:1411.7738 [hep-ex]} \BibitemShut {NoStop}%
\bibitem [{\citenamefont {Braaten}(2013)}]{Braaten:2013boa}%
  \BibitemOpen
  \bibfield  {author} {\bibinfo {author} {\bibfnamefont {E.}~\bibnamefont
  {Braaten}},\ }\href {https://doi.org/10.1103/PhysRevLett.111.162003}
  {\bibfield  {journal} {\bibinfo  {journal} {Phys. Rev. Lett.}\ }\textbf
  {\bibinfo {volume} {111}},\ \bibinfo {pages} {162003} (\bibinfo {year}
  {2013})},\ \Eprint {https://arxiv.org/abs/1305.6905} {arXiv:1305.6905
  [hep-ph]} \BibitemShut {NoStop}%
\bibitem [{\citenamefont {Dubynskiy}\ and\ \citenamefont
  {Voloshin}(2008)}]{Dubynskiy:2008mq}%
  \BibitemOpen
  \bibfield  {author} {\bibinfo {author} {\bibfnamefont {S.}~\bibnamefont
  {Dubynskiy}}\ and\ \bibinfo {author} {\bibfnamefont {M.~B.}\ \bibnamefont
  {Voloshin}},\ }\href {https://doi.org/10.1016/j.physletb.2008.07.086}
  {\bibfield  {journal} {\bibinfo  {journal} {Phys. Lett. B}\ }\textbf
  {\bibinfo {volume} {666}},\ \bibinfo {pages} {344} (\bibinfo {year}
  {2008})},\ \Eprint {https://arxiv.org/abs/0803.2224} {arXiv:0803.2224
  [hep-ph]} \BibitemShut {NoStop}%
\bibitem [{\citenamefont {De~Rujula}\ \emph {et~al.}(1977)\citenamefont
  {De~Rujula}, \citenamefont {Georgi},\ and\ \citenamefont
  {Glashow}}]{Rujula:PhysRevLett.38.317}%
  \BibitemOpen
  \bibfield  {author} {\bibinfo {author} {\bibfnamefont {A.}~\bibnamefont
  {De~Rujula}}, \bibinfo {author} {\bibfnamefont {H.}~\bibnamefont {Georgi}},\
  and\ \bibinfo {author} {\bibfnamefont {S.~L.}\ \bibnamefont {Glashow}},\
  }\href {https://doi.org/10.1103/PhysRevLett.38.317} {\bibfield  {journal}
  {\bibinfo  {journal} {Phys. Rev. Lett.}\ }\textbf {\bibinfo {volume} {38}},\
  \bibinfo {pages} {317} (\bibinfo {year} {1977})}\BibitemShut {NoStop}%
\bibitem [{\citenamefont {Tornqvist}(1994)}]{Tornqvist:ZPhysC61.525}%
  \BibitemOpen
  \bibfield  {author} {\bibinfo {author} {\bibfnamefont {N.~A.}\ \bibnamefont
  {Tornqvist}},\ }\href {https://doi.org/10.1007/BF01413192} {\bibfield
  {journal} {\bibinfo  {journal} {Z. Phys. C}\ }\textbf {\bibinfo {volume}
  {61}},\ \bibinfo {pages} {525} (\bibinfo {year} {1994})},\ \Eprint
  {https://arxiv.org/abs/hep-ph/9310247} {arXiv:hep-ph/9310247} \BibitemShut
  {NoStop}%
\bibitem [{\citenamefont {Anselmino}\ \emph {et~al.}(1993)\citenamefont
  {Anselmino}, \citenamefont {Predazzi}, \citenamefont {Ekelin}, \citenamefont
  {Fredriksson},\ and\ \citenamefont {Lichtenberg}}]{Anselmino:1992vg}%
  \BibitemOpen
  \bibfield  {author} {\bibinfo {author} {\bibfnamefont {M.}~\bibnamefont
  {Anselmino}}, \bibinfo {author} {\bibfnamefont {E.}~\bibnamefont {Predazzi}},
  \bibinfo {author} {\bibfnamefont {S.}~\bibnamefont {Ekelin}}, \bibinfo
  {author} {\bibfnamefont {S.}~\bibnamefont {Fredriksson}},\ and\ \bibinfo
  {author} {\bibfnamefont {D.~B.}\ \bibnamefont {Lichtenberg}},\ }\href
  {https://doi.org/10.1103/RevModPhys.65.1199} {\bibfield  {journal} {\bibinfo
  {journal} {Rev. Mod. Phys.}\ }\textbf {\bibinfo {volume} {65}},\ \bibinfo
  {pages} {1199} (\bibinfo {year} {1993})}\BibitemShut {NoStop}%
\bibitem [{\citenamefont {Jaffe}(2005)}]{Jaffe:2004ph}%
  \BibitemOpen
  \bibfield  {author} {\bibinfo {author} {\bibfnamefont {R.~L.}\ \bibnamefont
  {Jaffe}},\ }\href {https://doi.org/10.1016/j.physrep.2004.11.005} {\bibfield
  {journal} {\bibinfo  {journal} {Phys. Rept.}\ }\textbf {\bibinfo {volume}
  {409}},\ \bibinfo {pages} {1} (\bibinfo {year} {2005})},\ \Eprint
  {https://arxiv.org/abs/hep-ph/0409065} {arXiv:hep-ph/0409065} \BibitemShut
  {NoStop}%
\bibitem [{\citenamefont {Santopinto}\ and\ \citenamefont
  {Galata}(2007)}]{Santopinto:PhysRevC.75.045206}%
  \BibitemOpen
  \bibfield  {author} {\bibinfo {author} {\bibfnamefont {E.}~\bibnamefont
  {Santopinto}}\ and\ \bibinfo {author} {\bibfnamefont {G.}~\bibnamefont
  {Galata}},\ }\href {https://doi.org/10.1103/PhysRevC.75.045206} {\bibfield
  {journal} {\bibinfo  {journal} {Phys. Rev. C}\ }\textbf {\bibinfo {volume}
  {75}},\ \bibinfo {pages} {045206} (\bibinfo {year} {2007})},\ \Eprint
  {https://arxiv.org/abs/hep-ph/0605333} {arXiv:hep-ph/0605333} \BibitemShut
  {NoStop}%
\bibitem [{\citenamefont {Jaffe}(1999)}]{JAFFE_2001}%
  \BibitemOpen
  \bibfield  {author} {\bibinfo {author} {\bibfnamefont {R.~L.}\ \bibnamefont
  {Jaffe}},\ }in\ \href {https://doi.org/10.1142/9789812811530_0011} {\emph
  {\bibinfo {booktitle} {{Gregory Breit Centennial Symposium}}}}\ (\bibinfo
  {year} {1999})\ \Eprint {https://arxiv.org/abs/hep-ph/0001123}
  {arXiv:hep-ph/0001123} \BibitemShut {NoStop}%
\bibitem [{\citenamefont {Olsen}(2015{\natexlab{b}})}]{Olsen:2015zcy}%
  \BibitemOpen
  \bibfield  {author} {\bibinfo {author} {\bibfnamefont {S.~L.}\ \bibnamefont
  {Olsen}},\ }\href {https://doi.org/10.22323/1.238.0050} {\bibfield  {journal}
  {\bibinfo  {journal} {PoS}\ }\textbf {\bibinfo {volume} {Bormio2015}},\
  \bibinfo {pages} {050} (\bibinfo {year} {2015}{\natexlab{b}})},\ \Eprint
  {https://arxiv.org/abs/1511.01589} {arXiv:1511.01589 [hep-ex]} \BibitemShut
  {NoStop}%
\bibitem [{\citenamefont {Abe}\ \emph {et~al.}(2005{\natexlab{a}})\citenamefont
  {Abe} \emph {et~al.}}]{belle:abe2005evidence}%
  \BibitemOpen
  \bibfield  {author} {\bibinfo {author} {\bibfnamefont {K.}~\bibnamefont
  {Abe}} \emph {et~al.} (\bibinfo {collaboration} {Belle}),\ }in\ \href@noop {}
  {\emph {\bibinfo {booktitle} {{22nd International Symposium on Lepton-Photon
  Interactions at High Energy (LP 2005)}}}}\ (\bibinfo {year} {2005})\ \Eprint
  {https://arxiv.org/abs/hep-ex/0505037} {arXiv:hep-ex/0505037} \BibitemShut
  {NoStop}%
\bibitem [{\citenamefont {Abe}\ \emph {et~al.}(2005{\natexlab{b}})\citenamefont
  {Abe} \emph {et~al.}}]{belle:abe2005experimental}%
  \BibitemOpen
  \bibfield  {author} {\bibinfo {author} {\bibfnamefont {K.}~\bibnamefont
  {Abe}} \emph {et~al.} (\bibinfo {collaboration} {Belle}),\ }in\ \href@noop {}
  {\emph {\bibinfo {booktitle} {{22nd International Symposium on Lepton-Photon
  Interactions at High Energy (LP 2005)}}}}\ (\bibinfo {year} {2005})\ \Eprint
  {https://arxiv.org/abs/hep-ex/0505038} {arXiv:hep-ex/0505038} \BibitemShut
  {NoStop}%
\bibitem [{\citenamefont {Aaij}\ \emph {et~al.}(2015)\citenamefont {Aaij} \emph
  {et~al.}}]{LHCb:X3872PhysRevD.92.011102}%
  \BibitemOpen
  \bibfield  {author} {\bibinfo {author} {\bibfnamefont {R.}~\bibnamefont
  {Aaij}} \emph {et~al.} (\bibinfo {collaboration} {LHCb}),\ }\href
  {https://doi.org/10.1103/PhysRevD.92.011102} {\bibfield  {journal} {\bibinfo
  {journal} {Phys. Rev. D}\ }\textbf {\bibinfo {volume} {92}},\ \bibinfo
  {pages} {011102} (\bibinfo {year} {2015})},\ \Eprint
  {https://arxiv.org/abs/1504.06339} {arXiv:1504.06339 [hep-ex]} \BibitemShut
  {NoStop}%
\bibitem [{\citenamefont {Barnes}\ and\ \citenamefont
  {Godfrey}(2004)}]{Barnes:PhysRevD.69.054008}%
  \BibitemOpen
  \bibfield  {author} {\bibinfo {author} {\bibfnamefont {T.}~\bibnamefont
  {Barnes}}\ and\ \bibinfo {author} {\bibfnamefont {S.}~\bibnamefont
  {Godfrey}},\ }\href {https://doi.org/10.1103/PhysRevD.69.054008} {\bibfield
  {journal} {\bibinfo  {journal} {Phys. Rev. D}\ }\textbf {\bibinfo {volume}
  {69}},\ \bibinfo {pages} {054008} (\bibinfo {year} {2004})},\ \Eprint
  {https://arxiv.org/abs/hep-ph/0311162} {arXiv:hep-ph/0311162} \BibitemShut
  {NoStop}%
\bibitem [{\citenamefont {Tornqvist}(2004)}]{Tornqvist:x3872molecule}%
  \BibitemOpen
  \bibfield  {author} {\bibinfo {author} {\bibfnamefont {N.~A.}\ \bibnamefont
  {Tornqvist}},\ }\href {https://doi.org/10.1016/j.physletb.2004.03.077}
  {\bibfield  {journal} {\bibinfo  {journal} {Phys. Lett. B}\ }\textbf
  {\bibinfo {volume} {590}},\ \bibinfo {pages} {209} (\bibinfo {year}
  {2004})},\ \Eprint {https://arxiv.org/abs/hep-ph/0402237}
  {arXiv:hep-ph/0402237} \BibitemShut {NoStop}%
\bibitem [{\citenamefont {Swanson}(2004)}]{Swanson:x3872molecule}%
  \BibitemOpen
  \bibfield  {author} {\bibinfo {author} {\bibfnamefont {E.~S.}\ \bibnamefont
  {Swanson}},\ }\href {https://doi.org/10.1016/j.physletb.2004.07.059}
  {\bibfield  {journal} {\bibinfo  {journal} {Phys. Lett. B}\ }\textbf
  {\bibinfo {volume} {598}},\ \bibinfo {pages} {197} (\bibinfo {year}
  {2004})},\ \Eprint {https://arxiv.org/abs/hep-ph/0406080}
  {arXiv:hep-ph/0406080} \BibitemShut {NoStop}%
\bibitem [{\citenamefont {Choi}\ \emph {et~al.}(2008)\citenamefont {Choi} \emph
  {et~al.}}]{Belle:Z4430PhysRevLett.100.142001}%
  \BibitemOpen
  \bibfield  {author} {\bibinfo {author} {\bibfnamefont {S.~K.}\ \bibnamefont
  {Choi}} \emph {et~al.} (\bibinfo {collaboration} {Belle}),\ }\href
  {https://doi.org/10.1103/PhysRevLett.100.142001} {\bibfield  {journal}
  {\bibinfo  {journal} {Phys. Rev. Lett.}\ }\textbf {\bibinfo {volume} {100}},\
  \bibinfo {pages} {142001} (\bibinfo {year} {2008})},\ \Eprint
  {https://arxiv.org/abs/0708.1790} {arXiv:0708.1790 [hep-ex]} \BibitemShut
  {NoStop}%
\bibitem [{\citenamefont {Chilikin}\ \emph {et~al.}(2014)\citenamefont
  {Chilikin} \emph {et~al.}}]{Belle:Z4430PhysRevD.90.112009}%
  \BibitemOpen
  \bibfield  {author} {\bibinfo {author} {\bibfnamefont {K.}~\bibnamefont
  {Chilikin}} \emph {et~al.} (\bibinfo {collaboration} {Belle}),\ }\href
  {https://doi.org/10.1103/PhysRevD.90.112009} {\bibfield  {journal} {\bibinfo
  {journal} {Phys. Rev. D}\ }\textbf {\bibinfo {volume} {90}},\ \bibinfo
  {pages} {112009} (\bibinfo {year} {2014})},\ \Eprint
  {https://arxiv.org/abs/1408.6457} {arXiv:1408.6457 [hep-ex]} \BibitemShut
  {NoStop}%
\bibitem [{\citenamefont {Pascual}\ and\ \citenamefont
  {Tarrach}(1984)}]{Pascual:1984zb}%
  \BibitemOpen
  \bibfield  {author} {\bibinfo {author} {\bibfnamefont {P.}~\bibnamefont
  {Pascual}}\ and\ \bibinfo {author} {\bibfnamefont {R.}~\bibnamefont
  {Tarrach}},\ }\href@noop {} {\emph {\bibinfo {title} {QCD: Renormalization
  for the Practitioner}}},\ Vol.\ \bibinfo {volume} {194}\ (\bibinfo
  {publisher} {Springer-Verlag, Berlin},\ \bibinfo {year} {1984})\BibitemShut
  {NoStop}%
\bibitem [{\citenamefont {Peskin}\ and\ \citenamefont
  {Schroeder}(2018)}]{peskin}%
  \BibitemOpen
  \bibfield  {author} {\bibinfo {author} {\bibfnamefont {M.}~\bibnamefont
  {Peskin}}\ and\ \bibinfo {author} {\bibfnamefont {D.~V.}\ \bibnamefont
  {Schroeder}},\ }\href@noop {} {\emph {\bibinfo {title} {An Introduction To
  Quantum Field Theory}}}\ (\bibinfo  {publisher} {CRC Press},\ \bibinfo {year}
  {2018})\BibitemShut {NoStop}%
\bibitem [{\citenamefont {'t~Hooft}\ and\ \citenamefont
  {Veltman}(1972)}]{tHooft:1972tcz}%
  \BibitemOpen
  \bibfield  {author} {\bibinfo {author} {\bibfnamefont {G.}~\bibnamefont
  {'t~Hooft}}\ and\ \bibinfo {author} {\bibfnamefont {M.~J.~G.}\ \bibnamefont
  {Veltman}},\ }\href {https://doi.org/10.1016/0550-3213(72)90279-9} {\bibfield
   {journal} {\bibinfo  {journal} {Nucl. Phys. B}\ }\textbf {\bibinfo {volume}
  {44}},\ \bibinfo {pages} {189} (\bibinfo {year} {1972})}\BibitemShut
  {NoStop}%
\bibitem [{\citenamefont {Narison}\ and\ \citenamefont
  {Tarrach}(1983)}]{Narison:1983kn}%
  \BibitemOpen
  \bibfield  {author} {\bibinfo {author} {\bibfnamefont {S.}~\bibnamefont
  {Narison}}\ and\ \bibinfo {author} {\bibfnamefont {R.}~\bibnamefont
  {Tarrach}},\ }\href
  {https://doi.org/https://doi.org/10.1016/0370-2693(83)91271-6} {\bibfield
  {journal} {\bibinfo  {journal} {Physics Letters B}\ }\textbf {\bibinfo
  {volume} {125}},\ \bibinfo {pages} {217} (\bibinfo {year}
  {1983})}\BibitemShut {NoStop}%
\bibitem [{\citenamefont {Muta}(1987)}]{muta}%
  \BibitemOpen
  \bibfield  {author} {\bibinfo {author} {\bibfnamefont {T.}~\bibnamefont
  {Muta}},\ }\href {https://doi.org/10.1142/0022} {\emph {\bibinfo {title}
  {{Foundations of Quantum Chromodynamics: An Introduction to Perturbative
  Methods in Gauge Theories}}}},\ \bibinfo {edition} {3rd}\ ed.,\ \bibinfo
  {series} {World scientific Lecture Notes in Physics}, Vol.~\bibinfo {volume}
  {78}\ (\bibinfo  {publisher} {World Scientific},\ \bibinfo {year} {1987})\
  \Eprint
  {https://arxiv.org/abs/https://www.worldscientific.com/doi/pdf/10.1142/0022}
  {https://www.worldscientific.com/doi/pdf/10.1142/0022} \BibitemShut {NoStop}%
\bibitem [{\citenamefont {Gell-Mann}\ and\ \citenamefont
  {Low}(1954)}]{Gellmann:PhysRev.95.1300}%
  \BibitemOpen
  \bibfield  {author} {\bibinfo {author} {\bibfnamefont {M.}~\bibnamefont
  {Gell-Mann}}\ and\ \bibinfo {author} {\bibfnamefont {F.~E.}\ \bibnamefont
  {Low}},\ }\href {https://doi.org/10.1103/PhysRev.95.1300} {\bibfield
  {journal} {\bibinfo  {journal} {Phys. Rev.}\ }\textbf {\bibinfo {volume}
  {95}},\ \bibinfo {pages} {1300} (\bibinfo {year} {1954})}\BibitemShut
  {NoStop}%
\bibitem [{\citenamefont {Callan}(1970)}]{Callan:PhysRevD.2.1541}%
  \BibitemOpen
  \bibfield  {author} {\bibinfo {author} {\bibfnamefont {C.~G.}\ \bibnamefont
  {Callan}},\ }\href {https://doi.org/10.1103/PhysRevD.2.1541} {\bibfield
  {journal} {\bibinfo  {journal} {Phys. Rev. D}\ }\textbf {\bibinfo {volume}
  {2}},\ \bibinfo {pages} {1541} (\bibinfo {year} {1970})}\BibitemShut
  {NoStop}%
\bibitem [{\citenamefont {Symanzik}(1970)}]{Symanzik:CommMathPhys18.227}%
  \BibitemOpen
  \bibfield  {author} {\bibinfo {author} {\bibfnamefont {K.}~\bibnamefont
  {Symanzik}},\ }\href {https://doi.org/10.1007/BF01649434} {\bibfield
  {journal} {\bibinfo  {journal} {Commun. Math. Phys.}\ }\textbf {\bibinfo
  {volume} {18}},\ \bibinfo {pages} {227} (\bibinfo {year} {1970})}\BibitemShut
  {NoStop}%
\bibitem [{\citenamefont {Wilson}(1971)}]{Wilson:PhysRevD.3.1818}%
  \BibitemOpen
  \bibfield  {author} {\bibinfo {author} {\bibfnamefont {K.~G.}\ \bibnamefont
  {Wilson}},\ }\href {https://doi.org/10.1103/PhysRevD.3.1818} {\bibfield
  {journal} {\bibinfo  {journal} {Phys. Rev. D}\ }\textbf {\bibinfo {volume}
  {3}},\ \bibinfo {pages} {1818} (\bibinfo {year} {1971})}\BibitemShut
  {NoStop}%
\bibitem [{\citenamefont {Gross}\ and\ \citenamefont
  {Wilczek}(1973)}]{Gross:PhysRevLett.30.1343}%
  \BibitemOpen
  \bibfield  {author} {\bibinfo {author} {\bibfnamefont {D.~J.}\ \bibnamefont
  {Gross}}\ and\ \bibinfo {author} {\bibfnamefont {F.}~\bibnamefont
  {Wilczek}},\ }\href {https://doi.org/10.1103/PhysRevLett.30.1343} {\bibfield
  {journal} {\bibinfo  {journal} {Phys. Rev. Lett.}\ }\textbf {\bibinfo
  {volume} {30}},\ \bibinfo {pages} {1343} (\bibinfo {year}
  {1973})}\BibitemShut {NoStop}%
\bibitem [{\citenamefont {Politzer}(1973)}]{Politzer:PhysRevLett.30.1346}%
  \BibitemOpen
  \bibfield  {author} {\bibinfo {author} {\bibfnamefont {H.~D.}\ \bibnamefont
  {Politzer}},\ }\href {https://doi.org/10.1103/PhysRevLett.30.1346} {\bibfield
   {journal} {\bibinfo  {journal} {Phys. Rev. Lett.}\ }\textbf {\bibinfo
  {volume} {30}},\ \bibinfo {pages} {1346} (\bibinfo {year}
  {1973})}\BibitemShut {NoStop}%
\bibitem [{\citenamefont {Shifman}\ \emph
  {et~al.}(1979{\natexlab{a}})\citenamefont {Shifman}, \citenamefont
  {Vainshtein},\ and\ \citenamefont {Zakharov}}]{Shifman:1978bx}%
  \BibitemOpen
  \bibfield  {author} {\bibinfo {author} {\bibfnamefont {M.}~\bibnamefont
  {Shifman}}, \bibinfo {author} {\bibfnamefont {A.}~\bibnamefont
  {Vainshtein}},\ and\ \bibinfo {author} {\bibfnamefont {V.}~\bibnamefont
  {Zakharov}},\ }\href
  {https://doi.org/https://doi.org/10.1016/0550-3213(79)90022-1} {\bibfield
  {journal} {\bibinfo  {journal} {Nuclear Physics B}\ }\textbf {\bibinfo
  {volume} {147}},\ \bibinfo {pages} {385} (\bibinfo {year}
  {1979}{\natexlab{a}})}\BibitemShut {NoStop}%
\bibitem [{\citenamefont {Shifman}\ \emph
  {et~al.}(1979{\natexlab{b}})\citenamefont {Shifman}, \citenamefont
  {Vainshtein},\ and\ \citenamefont {Zakharov}}]{Shifman:1978by}%
  \BibitemOpen
  \bibfield  {author} {\bibinfo {author} {\bibfnamefont {M.}~\bibnamefont
  {Shifman}}, \bibinfo {author} {\bibfnamefont {A.}~\bibnamefont
  {Vainshtein}},\ and\ \bibinfo {author} {\bibfnamefont {V.}~\bibnamefont
  {Zakharov}},\ }\href
  {https://doi.org/https://doi.org/10.1016/0550-3213(79)90023-3} {\bibfield
  {journal} {\bibinfo  {journal} {Nuclear Physics B}\ }\textbf {\bibinfo
  {volume} {147}},\ \bibinfo {pages} {448} (\bibinfo {year}
  {1979}{\natexlab{b}})}\BibitemShut {NoStop}%
\bibitem [{\citenamefont {Reinders}\ \emph {et~al.}(1985)\citenamefont
  {Reinders}, \citenamefont {Rubinstein},\ and\ \citenamefont
  {Yazaki}}]{Reinders:1984sr}%
  \BibitemOpen
  \bibfield  {author} {\bibinfo {author} {\bibfnamefont {L.}~\bibnamefont
  {Reinders}}, \bibinfo {author} {\bibfnamefont {H.}~\bibnamefont
  {Rubinstein}},\ and\ \bibinfo {author} {\bibfnamefont {S.}~\bibnamefont
  {Yazaki}},\ }\href
  {https://doi.org/https://doi.org/10.1016/0370-1573(85)90065-1} {\bibfield
  {journal} {\bibinfo  {journal} {Physics Reports}\ }\textbf {\bibinfo {volume}
  {127}},\ \bibinfo {pages} {1} (\bibinfo {year} {1985})}\BibitemShut {NoStop}%
\bibitem [{\citenamefont {de~Rafael}(1997)}]{deRafael:1997ea}%
  \BibitemOpen
  \bibfield  {author} {\bibinfo {author} {\bibfnamefont {E.}~\bibnamefont
  {de~Rafael}},\ }in\ \href@noop {} {\emph {\bibinfo {booktitle} {{Les Houches
  Summer School in Theoretical Physics, Session 68: Probing the Standard Model
  of Particle Interactions}}}}\ (\bibinfo {year} {1997})\ pp.\ \bibinfo {pages}
  {1171--1218},\ \Eprint {https://arxiv.org/abs/hep-ph/9802448}
  {arXiv:hep-ph/9802448} \BibitemShut {NoStop}%
\bibitem [{\citenamefont {Colangelo}\ and\ \citenamefont
  {Khodjamirian}(2000)}]{Colangelo:2000dp}%
  \BibitemOpen
  \bibfield  {author} {\bibinfo {author} {\bibfnamefont {P.}~\bibnamefont
  {Colangelo}}\ and\ \bibinfo {author} {\bibfnamefont {A.}~\bibnamefont
  {Khodjamirian}},\ }in\ \href {https://doi.org/10.1142/9789812810458_0033}
  {\emph {\bibinfo {booktitle} {At The Frontier of Particle Physics}}},\
  \bibinfo {editor} {edited by\ \bibinfo {editor} {\bibfnamefont
  {M.}~\bibnamefont {Shifman}}\ and\ \bibinfo {editor} {\bibfnamefont
  {B.}~\bibnamefont {Ioffe}}}\ (\bibinfo {year} {2000})\ pp.\ \bibinfo {pages}
  {1495--1576},\ \Eprint {https://arxiv.org/abs/hep-ph/0010175}
  {arXiv:hep-ph/0010175} \BibitemShut {NoStop}%
\bibitem [{\citenamefont {Gubler}\ and\ \citenamefont
  {Satow}(2019)}]{Gubler:2018ctz}%
  \BibitemOpen
  \bibfield  {author} {\bibinfo {author} {\bibfnamefont {P.}~\bibnamefont
  {Gubler}}\ and\ \bibinfo {author} {\bibfnamefont {D.}~\bibnamefont {Satow}},\
  }\href {https://doi.org/https://doi.org/10.1016/j.ppnp.2019.02.005}
  {\bibfield  {journal} {\bibinfo  {journal} {Progress in Particle and Nuclear
  Physics}\ }\textbf {\bibinfo {volume} {106}},\ \bibinfo {pages} {1} (\bibinfo
  {year} {2019})}\BibitemShut {NoStop}%
\bibitem [{\citenamefont {Wilson}(1969)}]{Wilson:PhysRev.179.1499}%
  \BibitemOpen
  \bibfield  {author} {\bibinfo {author} {\bibfnamefont {K.~G.}\ \bibnamefont
  {Wilson}},\ }\href {https://doi.org/10.1103/PhysRev.179.1499} {\bibfield
  {journal} {\bibinfo  {journal} {Phys. Rev.}\ }\textbf {\bibinfo {volume}
  {179}},\ \bibinfo {pages} {1499} (\bibinfo {year} {1969})}\BibitemShut
  {NoStop}%
\bibitem [{\citenamefont {Gell-Mann}\ \emph {et~al.}(1968)\citenamefont
  {Gell-Mann}, \citenamefont {Oakes},\ and\ \citenamefont
  {Renner}}]{Gell-Mann:1968hlm}%
  \BibitemOpen
  \bibfield  {author} {\bibinfo {author} {\bibfnamefont {M.}~\bibnamefont
  {Gell-Mann}}, \bibinfo {author} {\bibfnamefont {R.~J.}\ \bibnamefont
  {Oakes}},\ and\ \bibinfo {author} {\bibfnamefont {B.}~\bibnamefont
  {Renner}},\ }\href {https://doi.org/10.1103/PhysRev.175.2195} {\bibfield
  {journal} {\bibinfo  {journal} {Phys. Rev.}\ }\textbf {\bibinfo {volume}
  {175}},\ \bibinfo {pages} {2195} (\bibinfo {year} {1968})}\BibitemShut
  {NoStop}%
\bibitem [{\citenamefont {Elias}\ \emph {et~al.}(1988)\citenamefont {Elias},
  \citenamefont {Steele},\ and\ \citenamefont
  {Scadron}}]{Elias:PhysRevD.38.1584}%
  \BibitemOpen
  \bibfield  {author} {\bibinfo {author} {\bibfnamefont {V.}~\bibnamefont
  {Elias}}, \bibinfo {author} {\bibfnamefont {T.~G.}\ \bibnamefont {Steele}},\
  and\ \bibinfo {author} {\bibfnamefont {M.~D.}\ \bibnamefont {Scadron}},\
  }\href {https://doi.org/10.1103/PhysRevD.38.1584} {\bibfield  {journal}
  {\bibinfo  {journal} {Phys. Rev. D}\ }\textbf {\bibinfo {volume} {38}},\
  \bibinfo {pages} {1584} (\bibinfo {year} {1988})}\BibitemShut {NoStop}%
\bibitem [{\citenamefont {Bagan}\ \emph {et~al.}(1994)\citenamefont {Bagan},
  \citenamefont {Ahmady}, \citenamefont {Elias},\ and\ \citenamefont
  {Steele}}]{Bagan:1994pw}%
  \BibitemOpen
  \bibfield  {author} {\bibinfo {author} {\bibfnamefont {E.}~\bibnamefont
  {Bagan}}, \bibinfo {author} {\bibfnamefont {M.~R.}\ \bibnamefont {Ahmady}},
  \bibinfo {author} {\bibfnamefont {V.}~\bibnamefont {Elias}},\ and\ \bibinfo
  {author} {\bibfnamefont {T.~G.}\ \bibnamefont {Steele}},\ }\href
  {https://doi.org/10.1007/BF01641898} {\bibfield  {journal} {\bibinfo
  {journal} {Z. Phys. C}\ }\textbf {\bibinfo {volume} {61}},\ \bibinfo {pages}
  {157} (\bibinfo {year} {1994})}\BibitemShut {NoStop}%
\bibitem [{\citenamefont {Bertlmann}\ \emph {et~al.}(1985)\citenamefont
  {Bertlmann}, \citenamefont {Launer},\ and\ \citenamefont
  {de~Rafael}}]{Bertlmann:1984ih}%
  \BibitemOpen
  \bibfield  {author} {\bibinfo {author} {\bibfnamefont {R.~A.}\ \bibnamefont
  {Bertlmann}}, \bibinfo {author} {\bibfnamefont {G.}~\bibnamefont {Launer}},\
  and\ \bibinfo {author} {\bibfnamefont {E.}~\bibnamefont {de~Rafael}},\ }\href
  {https://doi.org/10.1016/0550-3213(85)90475-4} {\bibfield  {journal}
  {\bibinfo  {journal} {Nucl. Phys. B}\ }\textbf {\bibinfo {volume} {250}},\
  \bibinfo {pages} {61} (\bibinfo {year} {1985})}\BibitemShut {NoStop}%
\bibitem [{\citenamefont {Kleiv}\ \emph {et~al.}(2013)\citenamefont {Kleiv},
  \citenamefont {Steele}, \citenamefont {Zhang},\ and\ \citenamefont
  {Blokland}}]{Kleiv:2013dta}%
  \BibitemOpen
  \bibfield  {author} {\bibinfo {author} {\bibfnamefont {R.~T.}\ \bibnamefont
  {Kleiv}}, \bibinfo {author} {\bibfnamefont {T.~G.}\ \bibnamefont {Steele}},
  \bibinfo {author} {\bibfnamefont {A.}~\bibnamefont {Zhang}},\ and\ \bibinfo
  {author} {\bibfnamefont {I.}~\bibnamefont {Blokland}},\ }\href
  {https://doi.org/10.1103/PhysRevD.87.125018} {\bibfield  {journal} {\bibinfo
  {journal} {Phys. Rev. D}\ }\textbf {\bibinfo {volume} {87}},\ \bibinfo
  {pages} {125018} (\bibinfo {year} {2013})},\ \Eprint
  {https://arxiv.org/abs/1304.7816} {arXiv:1304.7816 [hep-ph]} \BibitemShut
  {NoStop}%
\bibitem [{\citenamefont {Benmerrouche}\ \emph {et~al.}(1995)\citenamefont
  {Benmerrouche}, \citenamefont {Orlandini},\ and\ \citenamefont
  {Steele}}]{Benmerrouche:1995qa}%
  \BibitemOpen
  \bibfield  {author} {\bibinfo {author} {\bibfnamefont {M.}~\bibnamefont
  {Benmerrouche}}, \bibinfo {author} {\bibfnamefont {G.}~\bibnamefont
  {Orlandini}},\ and\ \bibinfo {author} {\bibfnamefont {T.}~\bibnamefont
  {Steele}},\ }\href
  {https://doi.org/https://doi.org/10.1016/0370-2693(95)00875-L} {\bibfield
  {journal} {\bibinfo  {journal} {Physics Letters B}\ }\textbf {\bibinfo
  {volume} {356}},\ \bibinfo {pages} {573} (\bibinfo {year}
  {1995})}\BibitemShut {NoStop}%
\bibitem [{\citenamefont {Jin}\ and\ \citenamefont
  {K\"orner}(2001)}]{Jin:2000ek}%
  \BibitemOpen
  \bibfield  {author} {\bibinfo {author} {\bibfnamefont {H.~Y.}\ \bibnamefont
  {Jin}}\ and\ \bibinfo {author} {\bibfnamefont {J.~G.}\ \bibnamefont
  {K\"orner}},\ }\href {https://doi.org/10.1103/PhysRevD.64.074002} {\bibfield
  {journal} {\bibinfo  {journal} {Phys. Rev. D}\ }\textbf {\bibinfo {volume}
  {64}},\ \bibinfo {pages} {074002} (\bibinfo {year} {2001})}\BibitemShut
  {NoStop}%
\bibitem [{\citenamefont {de~Oliveira}\ \emph {et~al.}(2022)\citenamefont
  {de~Oliveira}, \citenamefont {Harnett}, \citenamefont {Palameta},\ and\
  \citenamefont {Steele}}]{deOliveira:2022eeq}%
  \BibitemOpen
  \bibfield  {author} {\bibinfo {author} {\bibfnamefont {T.}~\bibnamefont
  {de~Oliveira}}, \bibinfo {author} {\bibfnamefont {D.}~\bibnamefont
  {Harnett}}, \bibinfo {author} {\bibfnamefont {A.}~\bibnamefont {Palameta}},\
  and\ \bibinfo {author} {\bibfnamefont {T.~G.}\ \bibnamefont {Steele}},\
  }\href {https://doi.org/10.1103/PhysRevD.106.114023} {\bibfield  {journal}
  {\bibinfo  {journal} {Phys. Rev. D}\ }\textbf {\bibinfo {volume} {106}},\
  \bibinfo {pages} {114023} (\bibinfo {year} {2022})},\ \Eprint
  {https://arxiv.org/abs/2208.12363} {arXiv:2208.12363 [hep-ph]} \BibitemShut
  {NoStop}%
\bibitem [{\citenamefont {Bogoliubov}\ and\ \citenamefont
  {Shirkov}(1980)}]{bogoliubov}%
  \BibitemOpen
  \bibfield  {author} {\bibinfo {author} {\bibfnamefont {N.~N.}\ \bibnamefont
  {Bogoliubov}}\ and\ \bibinfo {author} {\bibfnamefont {D.~V.}\ \bibnamefont
  {Shirkov}},\ }\href@noop {} {\emph {\bibinfo {title} {{Introduction to the
  Theory of Quantized Fields}}}}\ (\bibinfo  {publisher} {{Wiley, New York}},\
  \bibinfo {year} {1980})\BibitemShut {NoStop}%
\bibitem [{\citenamefont {Hepp}(1966)}]{Hepp:1966eg}%
  \BibitemOpen
  \bibfield  {author} {\bibinfo {author} {\bibfnamefont {K.}~\bibnamefont
  {Hepp}},\ }\href {https://doi.org/10.1007/BF01773358} {\bibfield  {journal}
  {\bibinfo  {journal} {Commun. Math. Phys.}\ }\textbf {\bibinfo {volume}
  {2}},\ \bibinfo {pages} {301} (\bibinfo {year} {1966})}\BibitemShut {NoStop}%
\bibitem [{\citenamefont {Zimmermann}(1969)}]{Zimmermann:1969jj}%
  \BibitemOpen
  \bibfield  {author} {\bibinfo {author} {\bibfnamefont {W.}~\bibnamefont
  {Zimmermann}},\ }\href {https://doi.org/10.1007/BF01645676} {\bibfield
  {journal} {\bibinfo  {journal} {Commun. Math. Phys.}\ }\textbf {\bibinfo
  {volume} {15}},\ \bibinfo {pages} {208} (\bibinfo {year} {1969})}\BibitemShut
  {NoStop}%
\bibitem [{\citenamefont {Collins}(1975)}]{Collins:1974da}%
  \BibitemOpen
  \bibfield  {author} {\bibinfo {author} {\bibfnamefont {J.~C.}\ \bibnamefont
  {Collins}},\ }\href {https://doi.org/10.1016/S0550-3213(75)80010-1}
  {\bibfield  {journal} {\bibinfo  {journal} {Nucl. Phys. B}\ }\textbf
  {\bibinfo {volume} {92}},\ \bibinfo {pages} {477} (\bibinfo {year}
  {1975})}\BibitemShut {NoStop}%
\bibitem [{\citenamefont {Collins}(1984)}]{Collins:1984xc}%
  \BibitemOpen
  \bibfield  {author} {\bibinfo {author} {\bibfnamefont {J.~C.}\ \bibnamefont
  {Collins}},\ }\href {https://doi.org/10.1017/CBO9780511622656} {\emph
  {\bibinfo {title} {Renormalization: An Introduction to Renormalization, the
  Renormalization Group and the Operator-Product Expansion}}},\ Cambridge
  Monographs on Mathematical Physics\ (\bibinfo  {publisher} {Cambridge
  University Press},\ \bibinfo {year} {1984})\BibitemShut {NoStop}%
\bibitem [{\citenamefont {Narison}(1988)}]{Narison:1988ep}%
  \BibitemOpen
  \bibfield  {author} {\bibinfo {author} {\bibfnamefont {S.}~\bibnamefont
  {Narison}},\ }\href {https://doi.org/10.1016/0370-2693(88)90379-6} {\bibfield
   {journal} {\bibinfo  {journal} {Phys. Lett. B}\ }\textbf {\bibinfo {volume}
  {210}},\ \bibinfo {pages} {238} (\bibinfo {year} {1988})}\BibitemShut
  {NoStop}%
\bibitem [{pri(2023)}]{private_communication}%
  \BibitemOpen
  \href@noop {} {\bibinfo {title} {In private communication with
  {R}.~{T}.~{K}leiv}} (\bibinfo {year} {2023})\BibitemShut {NoStop}%
\bibitem [{\citenamefont {Beringer}\ \emph {et~al.}(2012)\citenamefont
  {Beringer} \emph {et~al.}}]{ParticleDataGroup:2012pjm}%
  \BibitemOpen
  \bibfield  {author} {\bibinfo {author} {\bibfnamefont {J.}~\bibnamefont
  {Beringer}} \emph {et~al.} (\bibinfo {collaboration} {Particle Data Group}),\
  }\href {https://doi.org/10.1103/PhysRevD.86.010001} {\bibfield  {journal}
  {\bibinfo  {journal} {Phys. Rev. D}\ }\textbf {\bibinfo {volume} {86}},\
  \bibinfo {pages} {010001} (\bibinfo {year} {2012})}\BibitemShut {NoStop}%
\bibitem [{\citenamefont {Griffiths}(2008)}]{Griffiths:2008zz}%
  \BibitemOpen
  \bibfield  {author} {\bibinfo {author} {\bibfnamefont {D.}~\bibnamefont
  {Griffiths}},\ }\href@noop {} {\emph {\bibinfo {title} {{Introduction to
  elementary particles}}}}\ (\bibinfo  {publisher} {Wiley-vch},\ \bibinfo
  {year} {2008})\BibitemShut {NoStop}%
\end{thebibliography}%
\bibliographystyle{apsrev4-2}

\uofsappendix
\setstretch{1.5}

\chapter{Conventions}
\label{Appendix_A}
Contravariant four-vectors $x^{\mu}$ are defined as (see, e.g., Ref.~\cite{Griffiths:2008zz} for a review)
\begin{align}
	x^{\mu} = x = (x^0, x^1, x^2, x^3) = (ct, \vec x) \,.
\end{align}
Covariant four-vectors $x_{\mu}$ are defined as 
\begin{align}
	x_{\mu} \equiv g_{\mu\nu} x^{\nu} = (ct, -\vec x) \,,
\end{align}
where $g_{\mu\nu}$ is the metric matrix
\begin{align}
	g_{\mu\nu} = g^{\mu\nu} = \begin{pmatrix}
		1 & 0 & 0 & 0\\
		0 & -1 & 0 & 0 \\
		0 & 0 & -1 & 0 \\
		0 & 0 & 0 & -1
	\end{pmatrix} \,.
\end{align}
When the space-time dimension is $d$, $g_{\mu\nu}g^{\mu\nu}=d$.

Einstein's summation convention is assumed when there is a product with repeated indices, such as in the scalar product below
\begin{align}
	a.b \equiv a_{\mu}b^{\mu} = a_0b^0 + a_1b^1 + a_2b^2 + a_3b^3 = a^0b^0 - \vec a \cdot \vec b \,,
\end{align}
where the repeated index $\mu$ was summed from $\mu=0$ to $\mu=3$. In the gluon and quark colour space, assuming Einstein's summation convention, the product of delta functions becomes
\begin{subequations}
	\begin{align}
		&\delta^{ab}\delta_{ab} = \delta^a_a=N_c^2-1 \,, \\
		&\delta^{\alpha\beta}\delta_{\alpha\beta} = \delta^{\alpha}_{\alpha}=N_c \,,
	\end{align}
\end{subequations}
where $a$ is the gluon colour index, $\alpha$ is the quark colour index, and $N_c$ is the number of quark colours.

In natural units, $c$ and $\hbar$ are dimensionless constants where $c=1$ and $\hbar=1$. Equations, such as,
\begin{align}
	\label{eq:rel_energy}
	E^2=(\vec p c)^2 + (mc^2)^2
\end{align}
are then written as
\begin{align}
	E^2=(\vec p)^2 + m^2 \,.
	\label{eq:rel_energy_2}
\end{align}
Using dimensional analysis in Eq.~\eqref{eq:rel_energy_2} we have that energy, momentum, and mass have the same dimension: $[E]=[\vec p]=[m] = [\text{M}]^1$, where $[\text{M}]^1$ is a mass dimension equals to 1. The mass units broadly used in particle physics is the electron-volts (eV), where $1 \text{ eV}= 1.602 176 634\times10^{-19} \text{ J}$ \cite{pdg:2022}.

Since $[c] = [\text{L}]^1[\text{T}]^{-1}$ and $[\hbar] = [\text{M}]^1[\text{L}]^2[\text{T}]^{-1}$ in the International System of Units (SI units), where $[\text{L}]$ and $[\text{T}]$ are length and time dimensions respectively, in natural units $[c] = [\text{M}]^0$ and $[\hbar] = [\text{M}]^0$, therefore $[\text{M}]^1 = [\text{L}]^{-1} =[\text{T}]^{-1}$. Consequently, $[x^{\mu}]=[dt]=[\text{M}]^{-1}$, $[p^{\mu}]=[\partial^{\mu}]=[\text{M}]^1$, and $[d^d x]=[\text{M}]^{-d}$.

From the action equation
\begin{align}
	S=\int d^dx \mathcal{L}(x) \,,
\end{align}
where $[S]=[\hbar]=[\text{M}]^0$, the dimension for the Lagrangian density $\mathcal{L}(x)$ is $[\mathcal{L}(x)]=[\text{M}]^d$. 

Considering Eq.~\eqref{eq:qcd_lagrangian} we have
\begin{subequations}
	\begin{align}
		[\text{M}]^d &= [m_A\bar{q}^A_{\alpha}q^A_{\alpha}] \,, \ \ \therefore \ [\bar{q}^A_{\alpha}]=[q^A_{\alpha}] = [\text{M}]^{\frac{d-1}{2}} \,, \\ 
		[\text{M}]^d &= [\partial_{\mu}B^{\mu}_a(x)\ \partial_{\nu}B^{\nu}_a(x)] \,, \ \ \therefore \ [B^{\mu}_a(x)] = [\text{M}]^{\frac{d-2}{2}} \,, \\
		[\text{M}]^d &= [g^2\ B_{\mu}^b(x) B_{\nu}^c(x) B^{\mu}_d(x) B^{\nu}_e(x)] \,, \ \ \therefore \ [g] = [\text{M}]^{\frac{4-d}{2}}   \,.
	\end{align}   
\end{subequations}
Note that, in $d=4$, the QCD coupling $g$ is dimensionless.

The Dirac matrices $\gamma^{\mu}$ satisfy the following Dirac algebra and identity in $d$ dimensions \cite{Pascual:1984zb}
\begin{subequations}
	\begin{align}
		&\left\{\gamma^{\mu}, \gamma^{\nu}\right\} = 2g^{\mu\nu}, \ \ \mu, \nu = 0, 1, \dots, d-1 \,, \\
		&\gamma^{\mu}\gamma_{\mu} = d \,.
	\end{align}     
\end{subequations}
The gamma matrix $\gamma_{_5}$ is defined in $d=4$ as $\gamma_{_5} = i\gamma^{0}\gamma^{1}\gamma^{2}\gamma^{3}$, while in $d$ dimensions $\gamma_{_5}$ is defined in a way that
\begin{subequations}
	\begin{align}
		&\left\{\gamma_{_5}, \gamma_{\mu}\right\} = 0 \,, \\
		&(\gamma_{_5})^2 = 1 \,.
	\end{align}     
\end{subequations}

The Feynman slash notation $\slashed{a}$ is defined as $\slashed{a}=\gamma^{\mu} a_{\mu}$.

The charge conjugation operator $C$ appears in the diquark currents in Chapters \ref{chapter:Paper_1} and \ref{chapter:Paper_2}, and it is defined as
\begin{align}
	C = i\gamma^2\gamma^0 \,,
\end{align}
with the following identities
\begin{subequations}
	\begin{align}
		&C^{-1}=C^T=-C \,, \\
		&C^2=-1 \,, \\
		&C\slashed{a}^TC=\slashed{a} \,, \\
		&\left[C,\gamma_{_5}\right]=0 \,, \\
		&\gamma^0C=-C\gamma^0 \,,
	\end{align}
\end{subequations}
where $T$ denotes the transpose.

\end{document}